\documentclass[a4paper,12pt]{article}
\usepackage{epsfig,subfigure,amssymb,amscd,amsmath,graphicx,amsfonts,mathtools,mathcomp,pbox,amssymb,amsthm}
\usepackage[a4paper,top=20mm, bottom=20mm, left=20mm, right=20mm]{geometry}
\usepackage{times,math dots,hyperref,bbm}
\usepackage{tabularx}
\usepackage{authblk}
\usepackage[usenames]{color}
\usepackage{todonotes}

\usepackage[toc,page]{appendix}

\newtheorem{property}{Property}

\begin{document}

\title{Constructing edge zero modes through domain wall angle conservation}

\author[1]{D. Pellegrino\footnote{\url{domenico.pellegrino.2015@mumail.ie}}}
\author[2]{G. Kells\footnote{\url{gkells@stp.dias.ie}}}
\author[3]{N. Moran\footnote{\url{niall.moran@ichec.ie}}}
\author[1,2]{J. K. Slingerland\footnote{\url{joost@thphys.nuim.ie}}}
\affil[1]{\small Department of Theoretical Physics, Maynooth University, Ireland.}
\affil[2]{\small Dublin Institute for Advanced Studies, School of Theoretical Physics, 10 Burlington Rd, Dublin, Ireland}
\affil[3]{\small Irish Center for High-End Computing, National University of Ireland Galway, Ireland}

\date{\today} \maketitle
\begin{abstract}

\noindent We investigate the existence, normalization and explicit construction of edge zero modes in topologically ordered spin chains. In particular we give a detailed treatment of zero modes in a $\mathbb{Z}_3$ generalization of the Ising/Kitaev chain, which can also be described in terms of parafermions. We analyze when it is possible to iteratively construct strong zero modes, working completely in the spin picture. An important role is played by the so called total domain wall angle, a symmetry which appears in all models with strong zero modes that we are aware of. We show that preservation of this symmetry guarantees locality of the iterative construction, that is, it imposes locality conditions on the successive terms appearing in the zero mode's perturbative expansion. The method outlined here summarizes and generalizes some of the existing techniques used to construct zero modes in spin chains and sheds light on some surprising common features of all these types of methods. We conjecture a general algorithm for the perturbative construction of zero mode operators and test this on a variety of models, to the highest order we can manage. We also present analytical formulas for the zero modes which apply to all models investigated, but which feature a number of model dependent coefficients.

\end{abstract}

\section{Introduction}

Part of the interest in topological phases of matter stems from their potentially revolutionary technological applications.  One of the most exciting possibilities would be the ability to store and manipulate quantum information in topological degrees of freedom. This information would be intrinsically protected from some or all local error processes and there are now significant efforts being made to harness this property in scalable quantum devices \cite{Güll2018,Deng2016,Ruby2015,Nadj-Perge2014}.

The information in a topological quantum computation is typically stored in the system's ground state manifold, see e.g. \cite{Nayak2008}. Recently however there has been increased interest in whether the same topological properties can protect information at temperatures above the topological gap. In non-interacting topological superconductors \cite{Kitaev2001} such high-energy stability, naturally exists because the existence of a topological (Majorana) zero mode guarantees topological degeneracy at all energies.  A natural question then is if such high-temperature stability can exist in more realistic interacting systems.  

One of the main examples used to study this phenomenon is the so-called interacting p-wave wire/Kitaev chain. It is a well known fact that these models can be mapped to spin chains via the Jordan-Wigner transformation.  The presence of topological order in this model and similar ones, is signalled by the appearance of unpaired zero modes localised at the edges of the system when we take open boundary conditions \cite{Kitaev2001,Oreg2010}. In free fermion models, since the existence of these modes implies the presence of degeneracies throughout the whole spectrum, they are usually referred to as strong zero modes \cite{Fendley2012}; as opposed to weak (or almost strong) ones which would only act on a low-energy subspace and which give rise to degeneracy only within that subspace. The importance of the degeneracies high in the spectrum comes from the fact that this could possibly lead to high temperature fault-tolerant quantum computing \cite{Huse2013}.  In the interacting chain, strong zero modes have been established via bosonization arguments \cite{Gangadharaiah2011} and using iterative approaches at half filling, via a mapping to the XYZ chain \cite{XYZ}.  The Kitev chain is also known to have a region near the flat-band limit where interactions do not destroy the bulk topological degeneracy \cite{Kells2015, Kells2015b,Kemp_2017}.  Away from this regime there are indications that disorder induced localisation can help mitigate interaction-driven processes that destroy the topological degeneracy at high temperatures \cite{Huse2013,Kells2017}. 

More exotic interacting variants of the Kitaev chain are the so-called $\mathbb{Z}_N$ parafermionic clock models \cite{Fendley2012, Fendley2014}, which similarly to the Kitaev chain, also admit a description in terms of generalised spin chains. When written in these terms they are usually referred as chiral Potts models and the different types of phases that they possess have been subject of extensive studies \cite{Fradkin80,Howes1983,Gehlen1985,Ostlund1981,Ortiz2012}.
Notably there have been proposals for experimental realization of these parafermionic chains \cite{Lindner2012,Clarke2013,Alicea2016}. These models are also expected, in varying degrees, to possess high-temperature degeneracy, related to the presence of strong zero modes \cite{Jermyn2014,OurPaper,Else2017} (to some extent this is true also for the discrete group generalization of the clock models \cite{Munk2018}).  One of the most prominent regimes where this is expected to happen is around the chiral $\pi/6$ point of the $\mathbb{Z}_3$ model, where a constraint on total domain wall angle, prevents resonant decay processes that would otherwise destroy the topological stability at high energy densities. 

In this paper we analyse how this constraint on the domain wall angle \cite{OurPaper,Else2017} enters in the iterative construction of the associated zero-energy parafermionic modes.  In contrast to previous works we work entirely in the generalized spin picture and we provide a simple ansatz for the shape taken by the general solution of the problem, using the so called \textit{super-operator} formalism. In particular we show that the problem of finding zero modes at the edges of the system is related to a special type of degenerate perturbation theory of the \text{super-operator} Hamiltonian, which is obtained as the commutator with the original Hamiltonian, acting on the Hilbert space of operators. In this sense we are able to relate the domain wall symmetry to the locality of the terms appearing in this degenerate perturbation theory. Our approach can be readily extended to other types of spin chains and we find interesting common features between several of the studied models. In fact, we find that our treatment applies equally well to all spin models that we have considered.

We also address the problem of normalization of zero modes and we show how the properties of the zero mode for the unperturbed model induce similar behaviours upon the formal expansion of the zero mode for the perturbed one. Nonetheless, the problem regarding the existence of a finite radius of convergence for the formal perturbative series of the zero mode is still largely unanswered, but we are able to provide some encouraging results on this through numerical analysis. 

 In addition to general properties of strong zero modes, we also exhibit a method to construct the zero modes to any desired order of perturbation theory in the coupling constant of the interaction term in the Hamiltonian (in practice the order is limited only by computing resources). This method also generalizes straightforwardly to many spin models and we have used it extensively in guiding and checking our other results. We conjecture that the method is in fact an algorithm for finding zero modes to any order, but have not been able to prove this rigorously.  

An outline of the paper is as follows. In Section \ref{sec::TheModel} we review the relevant quantum clock model Hamiltonian and its properties which are relevant for the rest of the paper. Here we also introduce the concept of total domain-wall angle and how it relates to the super-operator picture and the iterative construction of the zero mode. In Sections \ref{sec::Symmetry} and \ref{sec::PerturbationTheory} we illustrate some relevant symmetries of the problem and then connect the iterative construction to the general framework of degenerate perturbation theory. In Section \ref{sec::LocalityTotalDomainWallParity} we illustrate how the total domain wall angle is related to the locality of the successive terms appearing in the perturbative expansion and in Section \ref{sec::GeneralFormOfTheSolution} we provide a general ansatz for the form taken by the solution of the formal expansion of the strong zero mode. In Section \ref{sec::Normalization} and \ref{sec::convergenceSeries} we consider the problem of normalization and convergence of the formal series. In section \ref{sec::AlgorithmicSolution} we describe a method (conjectured algorithm) for finding the zero mode, the results of which can be used to verify some of the claims made throughout the paper. Finally, in Section \ref{sec::Conclusions}, we draw our conclusions  and indicate possible directions for further research.

\section{The model}
\label{sec::TheModel}

\begin{figure}
\centering
\includegraphics[scale=0.6]{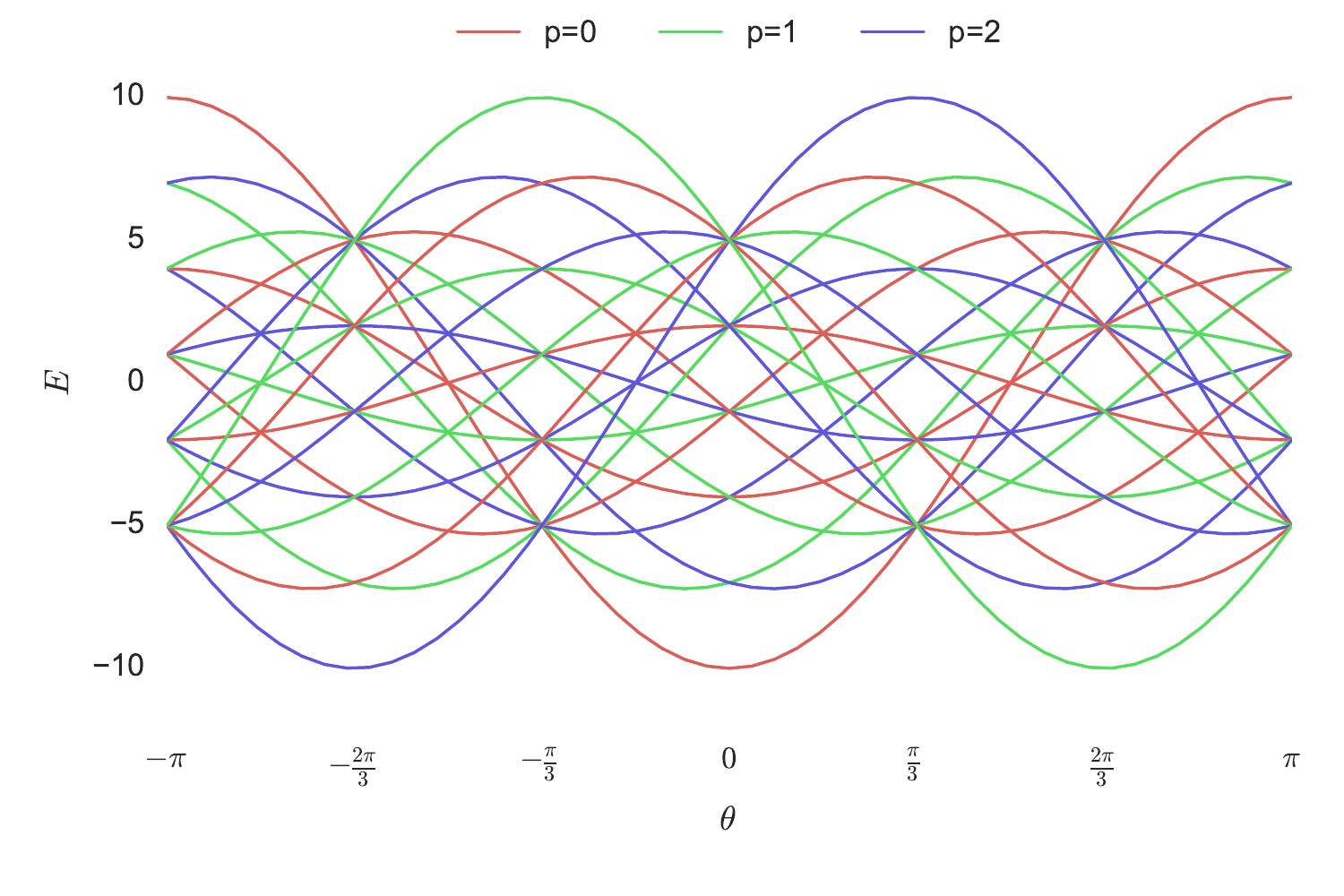}
\caption{Spectrum of $H_0$ for $L=6$. The different colors represent the different total domain wall angles.}
\label{fig::energyBands}
\end{figure}

We consider the $\mathbb{Z}_{3}$ quantum clock Hamiltonian as given in \cite{Clarke2013,Jermyn2014}, although our analysis can in principle be generalized to all $\mathbb{Z}_{N}$ with $N$ a prime number\footnote{When $N$ is not prime it is generally not possible to have strong zero modes, because of the presence of bands in the model that are everywhere degenerate \cite{OurPaper}.}. The Hamiltonian can be written in the form $H=H_0+fV$, where
\begin{equation}
\label{eq::InitialHamiltonian}
\begin{split}
&H_0=-\sum_{i=1}^{L-1}e^{i\theta}\sigma_i^\dagger\sigma_{i+1}+e^{-i\theta}\sigma_i\sigma_{i+1}^\dagger\\
&V=-\sum_{i=1}^Le^{i\phi}\tau_i+e^{-i\phi}\tau_i^\dagger
\end{split}
\end{equation}
with the convention
\begin{equation}
\sigma=\begin{pmatrix}1 & 0 & 0\\ 0 & \omega & 0\\ 0 & 0 & \omega^2\end{pmatrix}\qquad\tau=\begin{pmatrix}0 & 1 & 0\\ 0 & 0 & 1\\ 1 & 0 & 0\end{pmatrix}
\end{equation}
and $\omega=e^{\frac{2\pi i}{3}}$ is a third root of unity. The Hilbert space is spanned by vectors of the form
\begin{equation}
\label{eq::vectors}
|i_1,i_2,\ldots,i_L\rangle
\end{equation}
where $i_k \in \mathbb{Z}_3$ represents the values of the ``clocks'' at each site. This model can be viewed as a generalization of the Ising model ($N=2$). The commutation relations between $\sigma$ and $\tau$ matrices are 
\begin{equation}
\label{eq::commutationRelations}
\tau_{i}\sigma_{j}=\omega^{\delta_{i,j}}\sigma_j\tau_i\ .
\end{equation}
The following operator generalizes the fermion parity operator of the Ising model and plays a prominent role in the rest of the paper,
\begin{equation}
\label{eq::Q}
Q=\prod_{i=1}^L\tau_i\ .
\end{equation}
This operator moves the clock at each site one unit back. As $[H,Q]=0$, the spectrum of the Hamiltonian splits into three sectors, identified by the three eigenvalues of $Q$. In terms of the basis written above, $H_0$ is diagonal and its spectrum is given by
\begin{equation}
\label{eq::energyFreeModel}
E_{i_1,i_2,\ldots,i_L}=\sum_{k=1}^{L-1}\epsilon_{i_{k+1}-i_{k}}=n_0\epsilon_0+n_1\epsilon_1+n_2\epsilon_2
\end{equation}
where
\begin{equation}
\label{eq::domainWallEnergies}
\epsilon_{m}=-2\cos\left(\frac{2\pi m}{3}+\theta\right)
\end{equation}
and $n_m$ counts the number of domain walls of type $m$ in the state (\ref{eq::vectors}). We say that there is a domain wall of type $m$ between sites $k+1$ and $k$ when $i_{k+1}-i_{k}=m$. In particular,
the absence of domain wall is the same as a domain wall of type $0$. The typical energy bands of $H_0$ for different values of $\theta$ are shown in Fig. \ref{fig::energyBands}.
\\The energy of the unperturbed model is therefore determined by the set of vectors of type $(n_0,n_1,n_2)$ such that
\begin{equation}
n_0+n_1+n_2=L-1\ .
\end{equation}
Note that the spectrum of $H_0$ is the same in each $Q$-sector, so that three copies of each band exist, with different values of $Q$. This degeneracy may be split through the action of $V$. 

As shown in Ref.~\cite{OurPaper}, an important role is reserved for the $\theta$-values at which bands with different $(n_0,n_1,n_2)$ have the same energy.
These resonance points are further distinguished by another symmetry of $H_0$, namely the total domain wall angle, 
\[
P=\sigma_{1}^{\dagger}\sigma_{L}.
\]
Writing $P=e^{\frac{i 2\pi p}{3}}$, we get 
\begin{equation}
\label{eq::totalDomainWallAngle}
p=i_L-i_1 = \sum_{k=1}^{L-1}i_{k+1}-i_{k}=n_1+2n_2\quad \text{mod}\ 3\ ,
\end{equation}
so all the states in the $H_0$-band labeled by $(n_0,n_1,n_2)$ have total domain wall angle given by $p=n_1+2n_2$.
Strong zero modes may only occur at any given $\theta$ if either there are no bands crossing at this value of $\theta$ or, alternatively, if any bands that cross have the same value of $p$ (taken modulo 3). 
We note that there are in fact special values of $\theta$ at which the total domain wall angle is conserved even though different bands cross. One of these special points for the $N=3$ 
model is found at $\theta=\frac{\pi}{6}$, which also corresponds to the superintegrable point of the model when $\phi=\frac{\pi}{6}$ as well \cite{Howes1983,Gehlen1985} .  

\subsection{Parafermionic zero modes}\label{section::parafermionicZeroMode} 

Using a non-local transformation due to Fradkin and Kadanoff \cite{Fradkin80}, analogous to the Jordan--Wigner transformation, one can rewrite the spin model in terms of parafermionic variables. Parafermionic operators are defined as follows,
\[
\gamma_{2i-1}=\sigma_{i}\prod_{j<i}\tau_{j}\qquad\gamma_{2i}=\omega\sigma_{i}\prod_{j\le i}\tau_{j}\ 
\]
and satisfy the relations
\begin{equation}
\gamma_{i}\gamma_{j}=\omega^{\mathrm{sign}(j-i)}\gamma_{i}\gamma_{j}
\qquad\gamma_{i}^{N}=1\ .
\label{eq::paraf_alg}
\end{equation}
The Hamiltonian in terms of the parafermions basis is given by
\begin{equation}
\begin{split}
&H_{0}=-J\omega e^{i\theta}\sum_{i=1}^{L-1}\gamma^{\dagger}_{2i}\gamma^{\phantom{\dagger}}_{2i+1} + h.c.\\
&V=-f\omega e^{i\phi}\sum_{i=1}^{L}\gamma^{\dagger}_{2i-1}\gamma^{\phantom{\dagger}}_{2i} + h.c.
\end{split}
\end{equation}
and the parity operator becomes
\[Q=\omega^{-L}\prod_{i=1}^{L}\gamma^{\dagger}_{2i-1}\gamma^{\phantom{\dagger}}_{2i}
\]
In the exact same way as in the Ising case, there exist two parafermionic operators localized on the edges that commute with $H_0$, namely $\gamma_1$ and $\gamma_{2L}$. In terms of clock operators they are given by
\begin{equation}
\gamma_1=\sigma_1\qquad\gamma_{2L}=\omega\sigma_{L}Q\ .
\end{equation}
As in the Ising case a question arises over the existence of a localized zero mode when we introduce the potential $V$ term; but because of relation (\ref{eq::paraf_alg}) the situation is much more involved. 

The analysis of \cite{OurPaper} shows, through a perturbative approach, that $V$ can destroy the zero mode for values of $\theta$ where bands with different values of the total domain wall angle cross. The same perturbative analysis hints that the band degeneracy is left unbroken whenever the total domain wall angle is conserved.
Even though this persistence of the degeneracy strongly suggests 
the presence of zero modes whenever there are no bands crossing or when the crossing bands have the same total domain wall angle, there is no guarantee that \textit{localized} and \textit{normalizable} zero modes exist. Our aim is therefore to build up a recursive method, along the lines of \cite{Fendley2012,Munk2018}, that would allow the explicit construction of such operators. 

Considering the left edge of the system, it is reasonable to expect that if a localized zero mode $\psi$ exists, when we introduce the transverse field, this will reduce to $\gamma_1$ in the limit $f\rightarrow 0$. In other words we suppose that the zero mode admits the expansion

\begin{equation}
\label{eq::modeExpansion}
\psi=\psi^{(0)}+f\psi^{(1)}+\ldots+f^L\psi^{(L)}=\sum_{i=0}^Lf^i\psi^{(i)}\ ,
\end{equation} 

\noindent where $\psi^{(0)}=\gamma_1=\sigma_1$. The defining relations of the zero mode $\psi$ are similar to those satisfied by $\psi^{(0)}$
\begin{equation}
\label{eq::definingRelation1}
\left [H,\psi\right ]=O(e^{-\frac{L}{\xi}})
\end{equation}
with $1/\xi\propto\log(f)$ and 
\begin{equation}
\label{eq::definingRelation2}
Q\psi=\omega\psi Q
\end{equation}
which implies that the splitting between bands with the same $(n_0,n_1,n_2)$ in different $Q$ sectors vanishes exponentially with the length of the system. 

For the normalization we will mainly be concerned with the conditions
\begin{equation}
\label{eq::definingRelation31}
\psi^3=1+O(e^{-\frac{L}{\xi}})
\end{equation}
and 
\begin{equation}
    \psi^2=\psi^\dagger+O(e^{-\frac{L}{\xi}})\ .
\end{equation}
Together, they imply that the zero mode norm on a finite chain is given by  
\begin{equation}
\label{eq::definingRelation32}
\psi^{\dagger}\psi=1+O(e^{-\frac{L}{\xi}})\ .
\end{equation}
Despite this, the norm of $\psi$ may still diverge if the limit $L\rightarrow \infty$ is taken at fixed $f$, because the coefficient in front of the $f^{L}$ term may grow quickly with $L$, similarly to what happens in \cite{Kemp_2017} for spin chains. We will consider this further in Section \ref{sec::Normalization}. 

We stress that we will be interested in zero modes localised at the edge of the system. For finite systems, this property is ensured if matrix elements acting on sites that are at distance $l$ from the edge also appear at an order in the perturbing parameter $f$ that is equal or greater than $l$. However in the limit $L\mapsto \infty$ this may not be enough to guarantee that the mode is localised as the matrix elements themselves could grow faster than $f^L$, similarly to what happens to the normalization.

Let's now consider the commutator $[H,\psi]$. From (\ref{eq::modeExpansion}) we have
\begin{equation}
\label{eq::modeExpansionHamiltonian}
\begin{split}
&\left[H,\psi\right]=\sum_{k=1}^Lf^k\left(\left[H_0,\psi^{(k)}\right]+\left[V,\psi^{(k-1)}\right]\right)\ ,
\end{split}
\end{equation}
which means that in order for (\ref{eq::definingRelation1}) to be true (for all $f$) we must have
\begin{equation}
\label{eq::commutator}
\left[H_0,\psi^{(k)}\right]=-\left[V,\psi^{(k-1)}\right]\qquad k=1,2,\ \ldots,\ L-1\ .
\end{equation}
The commutators with $H_0$, and $V$ are linear operators acting on the Hilbert space of operators. Therefore, in principle, each of (\ref{eq::commutator}) gives a linear system that can be solved recursively once we fix $\psi^{(0)}$. In reality this method still presents certain ambiguities, related to the fact that the commutators with $H_0$, and $V$ are not invertible operators. We will address these in detail in the following.

To understand the point it is better to rewrite the operators $\mathcal{H}_0=\left[H_0,\cdot\ \right]$ and $\mathcal{V}=\left[V,\cdot\ \right]$ in a different form. The Hilbert space on which these commutators act is isomorphic to the Hilbert space for two copies of the original model. This can be established by using the so called Choi-Jamiolkowski isomorphism \cite{Choi1972,Jamiolkowski1972},
\begin{equation}
\label{eq::basis}
|i_1,i_2,\ldots ,i_L\rangle|j_1,j_2,\ldots ,j_L\rangle:=|i_1,i_2,\ldots ,i_L\rangle\langle j_1,j_2,\ldots ,j_L|
\end{equation}
with $|i_1,i_2,\ldots,i_L\rangle$ and $|j_1,j_2,\ldots,j_L\rangle$ as in (\ref{eq::vectors}). On each site this basis simply correspond to the canonical basis for matrices. For example we have
\begin{equation}
\begin{split}
   &\sigma=\begin{pmatrix}1&0&0\\0&\omega&0\\0&0&\omega^2\end{pmatrix}=|0\rangle|0\rangle+\omega|1\rangle|1\rangle+\omega^2|2\rangle|2\rangle\\
   &\tau=\begin{pmatrix}0&1&0\\0&0&1\\1&0&0\end{pmatrix}=|0\rangle|1\rangle+|1\rangle|2\rangle+|2\rangle|0\rangle\ .
\end{split}
\end{equation}
It can be easily seen that, given any operator $O$, in this basis, the commutator with $O$ takes the form
\begin{equation}
\label{eq::generalO}
    [O,\cdot\ ]=O\otimes \mathbbm{1}-\mathbbm{1}\otimes O^T
\end{equation}
where $O^T$ is the transpose of the operator $O$. In the following we will usually refer to these operators after their action on the Left and Right sectors of (\ref{eq::basis})
\begin{equation}
    O^L=O\otimes\mathbbm{1}\qquad O^R=\mathbbm{1}\otimes O
\end{equation}
Since $H_0$ and $V$ are hermitian and because of (\ref{eq::generalO}), we have
\begin{equation}
\begin{split}
&\mathcal{H}_0=[H_0,\cdot\ ]=H_0^L-{H_0^R}^T=H_0^L-{H_0^R}^*\\
&\mathcal{V}=[V,\cdot\ ]=V^L-{V^R}^T=V^L-{V^R}^*\ .
\end{split}
\end{equation}
Using a nomenclature coming from the study of open quantum systems, we will refer to operators acting on the space of operators as \emph{super-operators} \cite{Shaller2014}. Note that in the literature, the definition of right operators differs from our own by a transposition (that is $O^R=\mathbbm{1}\otimes O^T$).
\\The basis (\ref{eq::basis}) is precisely the one in which $\mathcal{H}_0$ is diagonal and we can write
\begin{equation}
 \begin{split}
&\mathcal{H}_0=\sum_{i=1}^{L-1}-e^{i\theta}{\sigma^L_{i+1}}^\dagger\sigma^L_i+e^{i\theta}{\sigma^R_{i+1}}^\dagger\sigma^R_i+h.c.\\
&\mathcal{V}=\sum_{i=1}^L-e^{i\phi}\tau^L_i+e^{-i\phi}\tau^R_i+h.c.\ .
\end{split}   
\end{equation}
It will be useful for the following to isolate the local structure of  $\mathcal{V}$, therefore we set
\begin{equation}
    \mathcal{V}_k=-e^{i\phi}\tau_k^L+e^{-i\phi}\tau_k^R+h.c.
\end{equation}
and $\mathcal{V}$ can be written as the sum of its local terms
\begin{equation}
    \mathcal{V}=\sum_{k=1}^{L}\mathcal{V}_k\ .
\end{equation}

Since $\mathcal{H}_0$ is diagonal we can easily find its eigenvalues, which we will indicate as
\begin{equation}
E_{j_1,j_2\ldots,j_L}^{i_1,i_2,\ldots,i_L}=E_{j_1,j_2,\ldots,j_L}-E_{i_1,i_2,\ldots,i_L}
\end{equation}   
with $E_{j_1,j_2,\ldots,j_L}$, $E_{i_1,i_2,\ldots,i_L}$ as in (\ref{eq::energyFreeModel}). The unperturbed energies of $H_0$ depend only on the domain walls along the chain. This means that when the left and right chain have the same domain walls at the endsite $L$, they will cancel out and we get
\begin{equation}
\label{eq::propertyEnergy}
E_{j_1,j_2\ldots,j_{L-1},i_L}^{i_1,i_2,\ldots,j_{L-1},i_L}=E_{j_1,j_2\ldots,j_{L-1}}^{i_1,i_2,\ldots,j_{L-1}}\qquad \forall\ i_L,j_{L-1}=0,1,2\ .
\end{equation}

Written in terms of the basis (\ref{eq::basis}) the operator $\psi^{(0)}$ takes the form
\begin{equation}
\label{eq::psi0}
\psi^{(0)}=\sigma_1=\sum_{i_1}\omega^{i_1}|i_1\rangle|i_1\rangle\otimes I_{L-1}
\end{equation}
where $I_k$ is the identity operator acting on a chain of length $k$:
\begin{equation}
\label{eq::identity}
I_{k}=\sum_{i_1,\ldots,i_{k}}|i_1,i_2,\ldots, i_{k}\rangle|i_1,i_2,\ldots, i_{k}\rangle\ .
\end{equation} 
In what follows, in order to simplify the notation we will often write the basis states in (\ref{eq::basis}) as 
$$|\mathbf{i}_k\rangle|\mathbf{j}_k\rangle\ ,$$
where $\mathbf{i}_k$, $\mathbf{j}_k$ represents the collection of the $k$ indices $i$s and $j$s. If $k<L$ it should be understood that there is an $I_{L-k}$ tensored at the end of the chain. For example we will write (\ref{eq::psi0}) as
\begin{equation}
\psi^{(0)}=\sum_{i}\omega^{i}|i\rangle|i\rangle
\end{equation}
In the cases where we need to highlight the relationship between specific indices we will often write expressions like
$$|\mathbf{i}_{t-1},i_t,\mathbf{i}_{k-t}\rangle|\mathbf{j}_{t-1},j_t,\mathbf{j}_{k-t}\rangle\ .$$

As a final remark we write down the form taken in the super-operator basis by the equations (\ref{eq::commutator}), which define the zero mode expansion:
\begin{equation}
    \label{eq::fundamentalEquation}
    \mathcal{H}_0\psi^{(k+1)}=-\mathcal{V}\psi^{(k)}\qquad k=1,2,\ldots,L-1\ .
\end{equation}

\section{Restrictions on the solutions}
\label{sec::Symmetry}

In this section we will consider super-operator symmetries of $\mathcal{H}_0$ and $\mathcal{V}$. These can be used to reduce the dimension of the space when we need to look for a solution of the problem. The first symmetry we will describe can be given in terms of the super-operator
\begin{equation}
\mathcal{Q}=\prod_{i=1}^L\tau^L_i\prod_{i=1}^L\tau^R_i.
\end{equation}
The use of $\mathcal{Q}$ allows to rewrite condition (\ref{eq::definingRelation2}) as \footnote{From our definitions we have that (\ref{eq::definingRelation2}) can be given as
\begin{equation*}
    Q^L\psi=\omega{Q^R}^T\psi\ .
\end{equation*}
Since in our basis $Q$ is orthogonal, equation (\ref{eq::definingRelation2SuperOperator}) follows.
}
\begin{equation}
\label{eq::definingRelation2SuperOperator}
\mathcal{Q}\psi=\omega\psi\ .
\end{equation}
Since $\mathcal{Q}\psi^{(0)}=\omega\psi^{(0)}$ and since $\mathcal{H}_0$ and $\mathcal{V}$ commute with $\mathcal{Q}$ we can impose this condition also at higher orders.  

The two super-operators $\mathcal{H}_0$ and $\mathcal{V}$ also have a chiral symmetry that reflects the commutator structure of their definition
\begin{equation}
    \begin{split}
        &(\mathcal{T}\mathcal{K})\mathcal{H}_0(\mathcal{T}\mathcal{K})^{-1}=-\mathcal{H}_0\\
        &(\mathcal{T}\mathcal{K})\mathcal{V}(\mathcal{T}\mathcal{K})^{-1}=-\mathcal{V}
    \end{split}
\end{equation}
where $\mathcal{T}$ is the super-operator which exchanges the left and right sectors,
\begin{equation}
\mathcal{T}|\mathbf{i}_L\rangle|\mathbf{j}_L\rangle=|\mathbf{j}_L\rangle|\mathbf{i}_L\rangle.
\end{equation}
and $\mathcal{K}$ is the complex conjugation
\begin{equation}
    \mathcal{K}(i\mathbbm{1})\mathcal{K}^{-1}=-i\mathbbm{1}\ .
\end{equation}
In other words $\mathcal{TK}$ simply corresponds to the conjugate transposition of operators.
As in the previous case, since 
\begin{equation}
    \mathcal{TK}(\psi^{(0)})=(\mathcal{K}\psi^{(0)})
\end{equation}
we can impose this condition, by induction, also at higher orders. By this we mean that the expansion obtained starting from $\mathcal{K}(\psi^{(0)})$, denoted as $(\mathcal{K}\psi)^{(k)}$, can be obtained as
\begin{equation}
\label{eq::conditionTK}
    (\mathcal{K}\psi)^{(k)}=\mathcal{TK}(\psi^{(k)})\ .
\end{equation}
Note that in the case $\phi=0$, both $\mathcal{H}_0$ and $\mathcal{V}$ are \textit{real} super-operators, and (\ref{eq::conditionTK}) reduces to
\begin{equation}
    \psi^{(k)}=\mathcal{T}\psi^{(k)}\ ,
\end{equation}
which can be used to diminish the dimensionality of the Hilbert space where we need to look for a solution.

\subsection{General considerations on $Null(\mathcal{H}_0)$}

Consider now a vector belonging to $Null(\mathcal{H}_0)$, which is non trivial only up to some site $k\leq L-1$. Explicitly, if the operator
\begin{equation}
|i_1,i_2,\ldots,i_{k-1},i_{k}\rangle|j_1,j_2,\ldots,j_{k-1},j_{k}\rangle\otimes I_{L-k}
\end{equation}
belongs to $Null(\mathcal{H}_0)$, because of (\ref{eq::propertyEnergy}), one must have
\begin{equation}
    \label{eq::energy&DomainWallAngle}
    E_{i_1,i_2,\ldots,i_k,l}=E_{j_1,j_2,\ldots,j_k,l}\qquad l=0,1,2\ .
\end{equation}
Concretely this means that an operator belong to $Null(\mathcal{H}_0)$ if it maps between states with the same band energy (with respect to $H_0$). 
Since $l$ is arbitrary, we can subtract two copies of the above equation with different values of the final clocks $l$ and $l'$. Using (\ref{eq::domainWallEnergies}) we get
\begin{equation*}
   E_{i_k,l}-E_{i_k,l'}=E_{j_k,l}-E_{j_k,l'}
\end{equation*}
and, after simple trigonometric manipulation, we end up with the condition
\begin{equation}
    \sin{\left((l+l'-2i_k)\frac{\pi}{3}+\theta\right)}=\sin{\left((l+l'-2j_k)\frac{\pi}{3}+\theta\right)}\ .
\end{equation}
Since $l$, $l'$ are arbitrary this equation can be true if and only if $i_k=j_k$. This means that operators that commute with $H_0$ cannot have $\tau$s on the last site where they act non-trivially. We remark that this has do be true independently of $\theta$.

Suppose now that $\theta$ is such that the total domain wall angle, as defined in (\ref{eq::totalDomainWallAngle}), is conserved. This means that 
\begin{equation}
    i_k-i_1=j_k-j_1\ .
\end{equation}
Since we just saw that for operators in $Null(\mathcal{H}_0)$ $i_k=j_k$, we  have
\begin{equation}
i_1=j_1\ . 
\end{equation}

\noindent We can therefore sum up these observations by the following statement:
\begin{property}
\label{property::Null}
A basis for the operators belonging to $Null(\mathcal{H}_0)$ and acting non trivially only up to some site $k$ is given by  
\begin{equation*}
    |\mathbf{i}_{k-1},i_k\rangle|\mathbf{j}_{k-1},i_{k}\rangle\qquad E^{\mathbf{i}_{k-1},i_k}_{\mathbf{j}_{k-1},i_{k}}=0
\end{equation*}
if, in addition, $\theta$ is such that the total domain wall angle is conserved, then the basis can be further restricted to
\begin{equation*}
    |i_1,\mathbf{i}_{k-2},i_k\rangle|i_1,\mathbf{j}_{k-1},i_{k}\rangle\qquad E^{i_1,\mathbf{i}_{k-1},i_k}_{i_1,\mathbf{j}_{k-1},i_{k}}=0\ .
\end{equation*}
\end{property}
\noindent As we will see this simple characterization of the vectors in $Null(\mathcal{H}_0)$ will be crucial when we will consider the existence of zero modes.

\section{Super-operators and perturbation theory} 
\label{sec::PerturbationTheory}
As seen in (\ref{eq::modeExpansionHamiltonian}) and (\ref{eq::commutator}), in order to find the zero mode, we need to solve the system of equations
\begin{equation}
\mathcal{H}_0\psi^{(k+1)}=-\mathcal{V}\psi^{(k)}\qquad k=1,\ldots,L-1\ .
\end{equation}
We will now show that this problem maps to a perturbation expansion of the super-Hamiltonian
\begin{equation}
\mathcal{H}=\mathcal{H}_0+f\mathcal{V}\ .
\end{equation}
Consider a $\psi$ as in (\ref{eq::modeExpansion}) such that
\begin{equation}
\label{eq::HE}
\mathcal{H}\psi=E\psi
\end{equation}
with 
\begin{equation}
    E=E^{(0)}+fE^{(1)}+f^2E^{(2)}+\ldots+f^LE^{(L)}
\end{equation}
If we call $\mathcal{P}_0$ the projector into $Null(\mathcal{H}_0)$ and $\mathcal{Q}_0=1-\mathcal{P}_0$, we can project (\ref{eq::HE}) into $Null(\mathcal{H}_0)$ and its orthogonal space. Using that $\mathcal{P}_0\mathcal{H}_0=0$ and $\mathcal{P}_0+\mathcal{Q}_0=1$ we have
\begin{equation}
\begin{split}
&f\mathcal{P}_0\mathcal{V}\psi=E\mathcal{P}_0\psi\\
&\mathcal{H}_0\mathcal{Q}_0\psi+f\mathcal{Q}_0\mathcal{V}\psi=E\mathcal{Q}_0\psi\ .
\end{split}
\end{equation}
Order by order this set of equations gives
\begin{equation}
\label{eq::BrillouinWignertemp}
\begin{split}
&\mathcal{P}_0\mathcal{V}\psi_p^{(k)}=-\mathcal{P}_0\mathcal{V}\psi_q^{(k)}+\sum_{i=0}^{k+1}E^{(k+1-i)}\psi_p^{(i)}\\
&\mathcal{H}_0\psi_q^{(k+1)}=-\mathcal{Q}_0\mathcal{V}\left(\psi_q^{(k)}+\psi_p^{(k)}\right)+\sum_{i=0}^{k+1} E^{(k+1-i)}\psi_q^{(i)}\ ,
\end{split}
\end{equation}
where $\psi_p^{(k)}=\mathcal{P}_0\psi^{(k)}$, $\psi_q^{(k)}=\mathcal{Q}_0\psi^{(k)}$ and these are the defining equations of any degenerate perturbation theory.
\\We can now see that finding the zero mode is equivalent to solve the perturbation theory problem when we impose  the conditions
\begin{equation}
\label{eq::energySplitting}
    E^{(j)}=0\qquad j\leq L-1\ ,
\end{equation}
which concretely means that we are asking our perturbation to not split the degeneracy up to order $L-1$. Equation (\ref{eq::BrillouinWignertemp}) thus reduces to 
\begin{align}
&\mathcal{P}_0\mathcal{V}\psi_p^{(k)}=-\mathcal{P}_0\mathcal{V}\psi_q^{(k)}\label{eq::BrillouinWigner1}\\
&\mathcal{H}_0\psi_q^{(k+1)}=-\mathcal{Q}_0\mathcal{V}\left(\psi_q^{(k)}+\psi_p^{(k)}\right)\label{eq::BrillouinWigner2}
\end{align}
which in turn is equivalent to (\ref{eq::fundamentalEquation}).

Usually in degenerate perturbation theory one looks for solutions that split the energy at some order. This is therefore a quite strange perturbation expansion, as we are specifically looking for a solution of the perturbative problem that does not split the energy at any order. Moreover we explicitly require that $\psi^{(0)}=\sigma_1$, while generally the starting point of the perturbation, inside the degenerate space, is to some extent undetermined. This makes the existence of a solution of (\ref{eq::BrillouinWigner1}) and (\ref{eq::BrillouinWigner2}) highly non trivial. In particular we know from \cite{OurPaper}, that if we are at resonance points where the total domain wall angle is not conserved, the system will develop energy splitting at order strictly less than $L$ and this means that a solution of (\ref{eq::BrillouinWigner1}, \ref{eq::BrillouinWigner2}) is generally not possible.

To understand the structure of the potential solution let's now consider the action of $\mathcal{H}_0$ and $\mathcal{V}$. The operator $\mathcal{V}$ can act at most on one element of the chains, while $\mathcal{H}_0$ can act on two elements. This means that, if as in (\ref{eq::psi0}) we start with an operator that is localized on the edge, and a solution of (\ref{eq::fundamentalEquation}) exists, the next order will be made of operators different from the identity for at most one site more than the previous order. For this reason, from now on we will consider each $\psi^{(k)}$ to be operators acting non-trivially on a chain of length $k+1$. In this sense (\ref{eq::fundamentalEquation}) can be rewritten as 
\begin{equation}
\label{eq::fundamentalEquationIdentity}
\mathcal{H}_0\psi^{(k+1)}=-\mathcal{V}\psi^{(k)}\otimes I_1
\end{equation} 
With this notation we  can rewrite (\ref{eq::BrillouinWigner1}, \ref{eq::BrillouinWigner2}) as
\begin{align}
    &\mathcal{P}_0\mathcal{V}(\psi_p^{(k)}\otimes I_1)=-\mathcal{P}_0\mathcal{V}(\psi_q^{(k)}\otimes I_1)\label{eq::BrillouinWigner1Identity}\\
    &\mathcal{H}_0\psi_q^{(k+1)}=-\mathcal{Q}_0\mathcal{V}\left(\psi_q^{(k)}+\psi_p^{(k)}\right)\otimes I_1\label{eq::BrillouinWigner2Identity}\ .
\end{align}
If a solution exists, equation (\ref{eq::BrillouinWigner1Identity}) implies that
\begin{equation}
\label{eq::conditionExistenceTemp}
\mathcal{V}\left(\psi_p^{(k)}+\psi_q^{(k)}\right)\otimes I_1\in Null(\mathcal{H}_0)^\perp\ .
\end{equation}
When this latter condition is satisfied we can easily find a solution for $\psi_q^{(k+1)}$ by inverting $\mathcal{H}_0$
\begin{equation}
\label{eq::chiqTemp}
\psi_q^{(k+1)}=-\frac{1}{\mathcal{H}_0}\mathcal{V}\left(\psi_p^{(k)}+\psi_q^{(k)}\right)\otimes I_1\ ,
\end{equation}
where we can omit the $\mathcal{Q}_0$ because of (\ref{eq::conditionExistenceTemp}). Hence, if we find a $\psi_p^{(k)}$ such that (\ref{eq::BrillouinWigner1Identity}) holds, we can always find a $\psi_q^{(k+1)}$ which makes the iterative construction work; and if $\psi^{(k)}_p$ acts on chains only up to order $k+1$ then the resulting zero mode will still be localized on the edge by construction. Note however that nothing prevents $\mathcal{P}_0\mathcal{V}\psi^{(k)}_q\otimes I_1$ from containing terms that live on chains of length $k+2$ (remember that $\psi_q^{(k)}$ lives on a chain of length $k+1$). We will see that if this is the case, then a ``local'' solution for $\psi_p^{(k)}$ will generally not exist (see Property \ref{property::NonExistence}). 

In this section we showed the importance of $\psi_p^{(k)}$ and equation (\ref{eq::chiqTemp}) shows that the problem of finding zero modes is essentially a problem of finding $\psi^{(k)}_p$. In the following we will carefully consider the situations in which such a ``local'' $\psi^{(k)}_p$ exists and when it does not.

\section{General considerations on $\psi_p$ }
\label{sec::RestrictionOnTheSolution}

Here we will consider what are the general properties of the solution $\psi_p^{(k)}$ of (\ref{eq::BrillouinWigner1Identity}). To this end we keep the discussion general by considering $\alpha_p,\ \beta_p\in Null(\mathcal{H}_0)$ such that
\begin{equation}
\label{eq::necessaryConditions}
    \mathcal{P}_0\mathcal{V}\alpha_p\otimes I_1=\beta_p\otimes I_1
\end{equation}
Later on we will specialize $\alpha_p$ and $\beta_p$ to $\psi_p^{(k)}$ and $\mathcal{P}_0\mathcal{V}(\psi_q^{(k)}\otimes I_1)$ respectively. 
\\Now take $\beta_p$ to be some operator that acts non trivially only up to some site $t$. From Property \ref{property::Null} we know that
\begin{equation}
\label{eq::necessaryConditionBeta}
    \beta_p=\sum\beta^{\mathbf{i}_{t-1},i_t}_{\mathbf{j}_{t-1},i_t}|\mathbf{i}_{t-1},i_t\rangle|\mathbf{j}_{t-1},i_t\rangle
\end{equation}
for some coefficients $\beta_{\mathbf{i}_{t-1},i_t}^{\mathbf{j}_{t-1},i_t}$. In the same way, $\alpha_p$ is an operator that acts non-trivially only up to some site $t'$ such that
\begin{equation}
\label{eq::restrictionChiP}
    \alpha_p=\sum\alpha^{\mathbf{i}_{t'-1},i_{t'}}_{\mathbf{j}_{t-1},i_{t'}}|\mathbf{i}_{t'-1},i_{t'}\rangle|\mathbf{j}_{t'-1},i_{t'}\rangle\ ,
\end{equation}
then following property then holds:
\begin{property}
\label{property::GeneralConditionAlphaBeta}
Consider the equation
\begin{equation*}
    \mathcal{P}_0\mathcal{V}\alpha_p\otimes I_1=\beta_p\otimes I_1
\end{equation*}
with $\alpha_p,\beta_p\in Null(\mathcal{H}_0)$ as in (\ref{eq::necessaryConditionBeta}) and (\ref{eq::restrictionChiP}). If we assume that this equation holds than necessarily $t'=t$ and
\begin{equation}
\label{eq::ConditionExistenceAlpha}
    \alpha_p=\sum\alpha^{\mathbf{i}_{t-2},i_{t-1}}_{\mathbf{j}_{t-2},i_{t-1}}|\mathbf{i}_{t-2},i_{t-1}\rangle|\mathbf{j}_{t-2},i_{t-1}\rangle\otimes I_1\ .
\end{equation}
\end{property}
\noindent This property means that, if $\beta_p$ acts non-trivially up to some site $t$, then equation (\ref{eq::conditionExistence}) imposes that $\alpha_p$ can act non-trivially only up to site $t-1$. The proof of this statement is given in Appendix \ref{appendix::ProofProperty24}. The idea behind it is that, since we have $I_1$ at the end of both $\alpha_p$ and $\beta_p$, then the action of $\mathcal{P}_0\mathcal{V}$ on $\alpha_p$ is inconsistent with the equality unless
\begin{equation}
    \mathcal{V}_t\alpha_p=0\ .
\end{equation}
As an example of this consider the state
\begin{equation*}
    |0,2,0\rangle|0,1,0\rangle\otimes I_1 ,
\end{equation*}
which belongs to $Null(\mathcal{H}_0)$ for a system of length $3$ and general $\theta$. When we act with $\mathcal{P}_0\mathcal{V}_3$ we obtain
\begin{equation*}
    \begin{split}
        \mathcal{P}_0\mathcal{V}_3(|0,2,0\rangle|0,1,0\rangle\otimes I_1)=&-e^{-i\phi}|0,2,1,2\rangle|0,1,0,2\rangle-e^{i\phi}|0,2,2,0\rangle|0,1,0,0\rangle\\
        &+e^{i\phi}|0,2,0,0\rangle|0,1,1,0\rangle+e^{-i\phi}|0,2,0,1\rangle|0,1,2,1\rangle\ .
    \end{split}
\end{equation*}
Similarly, whatever initial vector we pick on the left hand side, we will always find operators still belonging to $Null(\mathcal{H}_0)$ and whose left-right indices differ on the third site. In other words, the action of $\mathcal{P}_0\mathcal{V}_3$ is never trivial, unless we have an identity operator present on the third site. This fact, appropriately generalized, together with (\ref{eq::necessaryConditions}), implies Property \ref{property::GeneralConditionAlphaBeta} (see Appendix \ref{appendix::ProofProperty24} for more details).

In the context of finding the zero modes expansion this property means that if $\mathcal{P}_0\mathcal{V}(\psi_q^{(k)}\otimes I_1)$ contains operators that act up to chain of length $t$, then if $\psi_p^{(k)}$ exists, it has to acts on ``smaller'' chains. This can be used to reduce the dimension of the space where we need to look for a solution of the problem.

\subsection{Starting point of the expansion}
The previous discussion raises some new questions. Suppose that for a given $\theta$ a solution for the zero mode expansion that starts at $\psi^{(0)}=\sigma_1$ does not exist. It is reasonable to ask whether it is possible to fix $\psi^{(0)}\in Null(\mathcal{H}_0)$ differently, in a way that would allow a solution of the perturbative problem to exist. As a first step towards the answer to this question, we consider equations (\ref{eq::BrillouinWigner1Identity}, \ref{eq::BrillouinWigner2Identity}) at zeroth order
\begin{equation}
\label{eq::SolutionChip0}
    \mathcal{P}_0\mathcal{V}(\psi_p^{(0)}\otimes I_1)=0\qquad \psi_q^{(0)}=0
\end{equation}
from the proof of Property \ref{property::GeneralConditionAlphaBeta} we see that in order for this equation to have a solution we are forced to have
\begin{equation}
    \psi_p^{(0)}=\sum_{i_1}\psi^{i_1}_{i_1}|i_1\rangle|i_1\rangle\ .
\end{equation}
This means that a basis of solutions is determined by
\begin{equation}
    |0\rangle|0\rangle\qquad|1\rangle|1\rangle\qquad|2\rangle|2\rangle\ ,
\end{equation}
except when $\theta=0,\frac{\pi}{3},\frac{2\pi}{3}$, in which cases a solution such that $\mathcal{Q}\psi^{(0)}=\omega\psi^{(0)}$ does not exist\footnote{This can be easily seen by considering that at these resonance points there are additional states present in $Null(\mathcal{H}_0)$. For example consider $\theta=0$ and, to simplify the notation, $\phi=0$. We have
\begin{equation*}
\begin{split}
    &\mathcal{P}_0\mathcal{V}|0\rangle|0\rangle\otimes I_1=-|1,2\rangle|0,2\rangle-|2,1\rangle|0,1\rangle+|0,1\rangle|2,1\rangle+|0,2\rangle|1,2\rangle\\
    &\mathcal{P}_0\mathcal{V}|1\rangle|1\rangle\otimes I_1=-|0,2\rangle|1,2\rangle-|2,0\rangle|1,0\rangle+|1,2\rangle|0,2\rangle+|1,0\rangle|2,0\rangle\\
    &\mathcal{P}_0\mathcal{V}|2\rangle|2\rangle\otimes I_1=-|0,1\rangle|2,1\rangle-|1,0\rangle|2,0\rangle+|2,1\rangle|0,1\rangle+|2,0\rangle|1,0\rangle
\end{split}    
\end{equation*}
 It can be seen now that the unique solution of (\ref{eq::SolutionChip0}) that gives $0$ is $$I_1=|0\rangle|0\rangle+|1\rangle|1\rangle+|2\rangle|2\rangle\ ,$$ which is inconsistent with the condition $\mathcal{Q}\psi^{(0)}=\omega\psi^{(0)}$. A similar analysis can be conducted for the other $\theta$s and $\phi\neq0$.}. Summing up, we have restrictions on the possible local starting points for the zero mode expansion and the following property holds
\begin{property}
\label{property::Solution0}
When $\theta\neq0,\frac{\pi}{3},\frac{2\pi}{3}$, the unique possible local starting point of the zero mode expansion is given, up to a multiplicative constant, by
\begin{equation}
    \psi^{(0)}_p=\sigma_1\ .
\end{equation}
When $\theta=0,\frac{\pi}{3},\frac{2\pi}{3}$ a local zero mode cannot exists.
\end{property}

\section{Locality and total domain wall angle}
\label{sec::LocalityTotalDomainWallParity}

As we saw in the previous sections the question about the existence of an expansion for the zero mode is really a question about the existence of an expansion $\psi_p\in Null(\mathcal{H}_0)$ and we have explored to some extent what restrictions we can impose to the solution. In particular we have seen that the ``length'' of $\psi_p^{(k)}$ is necessarily smaller than the ``length'' of $\psi_q^{(k)}$. 

In this section we will concentrate on the other side of equation (\ref{eq::BrillouinWigner1Identity}), that is $\mathcal{P}_0\mathcal{V}(\psi_q\otimes I_1)$. In particular we will discuss the restrictions that the total domain wall angle imposes on it. In this sense the presence of total domain wall angle conservation translates to a condition imposed upon the locality of the terms appearing in the expansion of $\mathcal{P}_0(\mathcal{V}\psi_q^{(k)}\otimes I_1)$. The following property can be established:

\begin{property}
\label{property::conditionExistence}
Suppose that a local solution for the zero mode expansions exists up to some order $k$. If the total domain wall angle is conserved then
\begin{equation}
\label{eq::conditionExistence}
     \mathcal{P}_0(\mathcal{V}\psi_q^{(k)}\otimes I_1)=\beta_p^{(k)}\otimes I_1\ .
\end{equation}
with $\beta_p^{(k)}\in Null(\mathcal{H}_0)$ as in Property \ref{property::Null} and of the same ``length'' of $\psi^{(k)}_q$.
\end{property}
\noindent The proof of this statement is given in Appendix \ref{appendix:ProofProperty5}. Note however that this statement is not at all trivial. Consider for example the operator $|0,0\rangle|1,1\rangle$, which belongs to $Null(\mathcal{H}_0)$ for any $\theta$. Then
\begin{equation*}
    \mathcal{P}_0\mathcal{V}(|0,0\rangle|1,1\rangle\otimes I_1)=e^{i\phi}|0,0,1\rangle|0,1,1\rangle-e^{-i\phi}|1,0,0\rangle|1,1,0\rangle\ ,
\end{equation*}
which, evidently, cannot be written as $\beta_p\otimes I_1$ with $\beta_p$ of length 2. Concretely this property means that projecting the term $\mathcal{V}\psi_q^{(k)}$ to the null space of $\mathcal{H}_0$ does not result in any non-trivial elements further up the chain when total domain wall angle is conserved.
\begin{figure}[t]
\centering
  \subfigure[]{\includegraphics[scale=0.40]{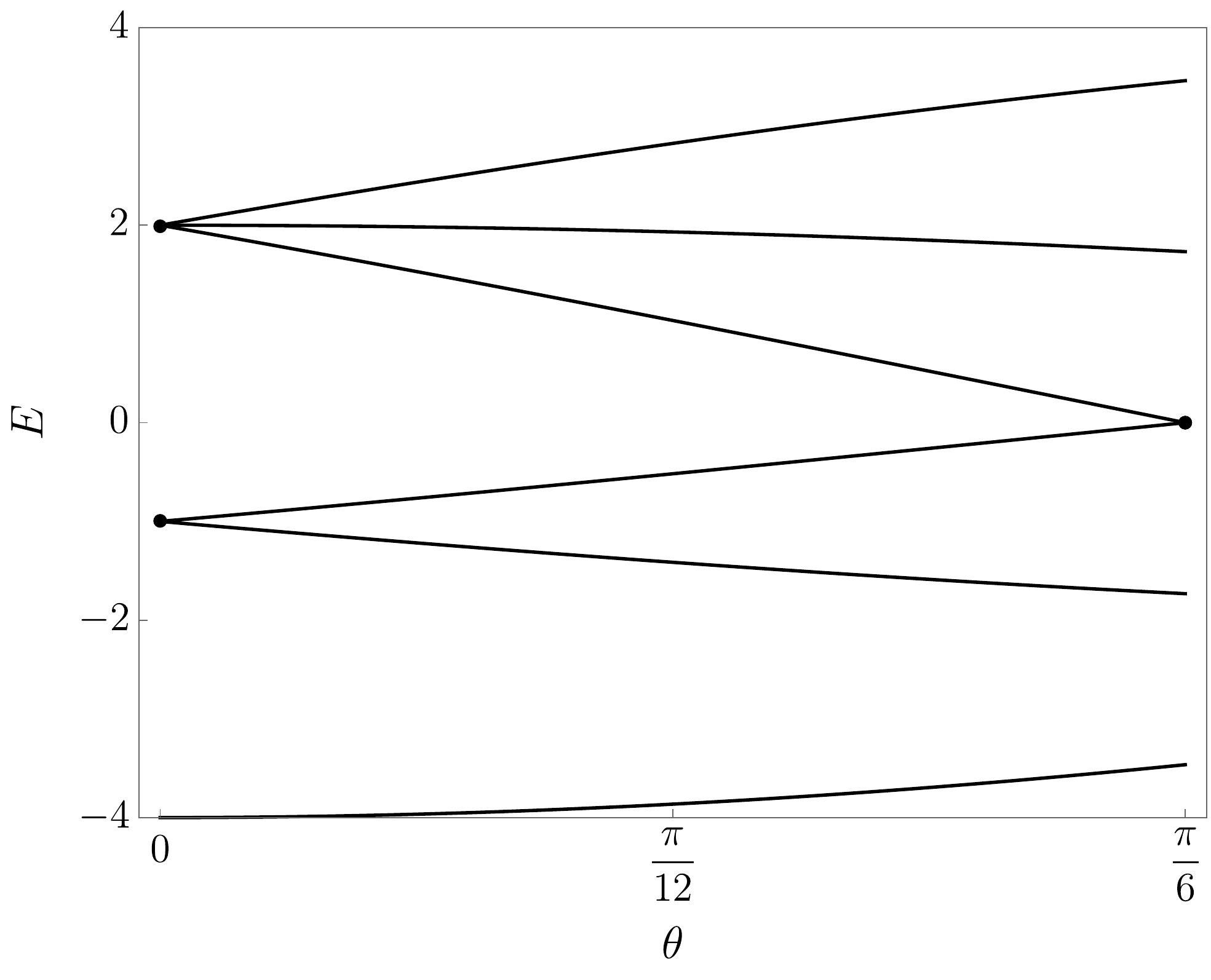}}\quad
  \subfigure[]{\includegraphics[scale=0.40]{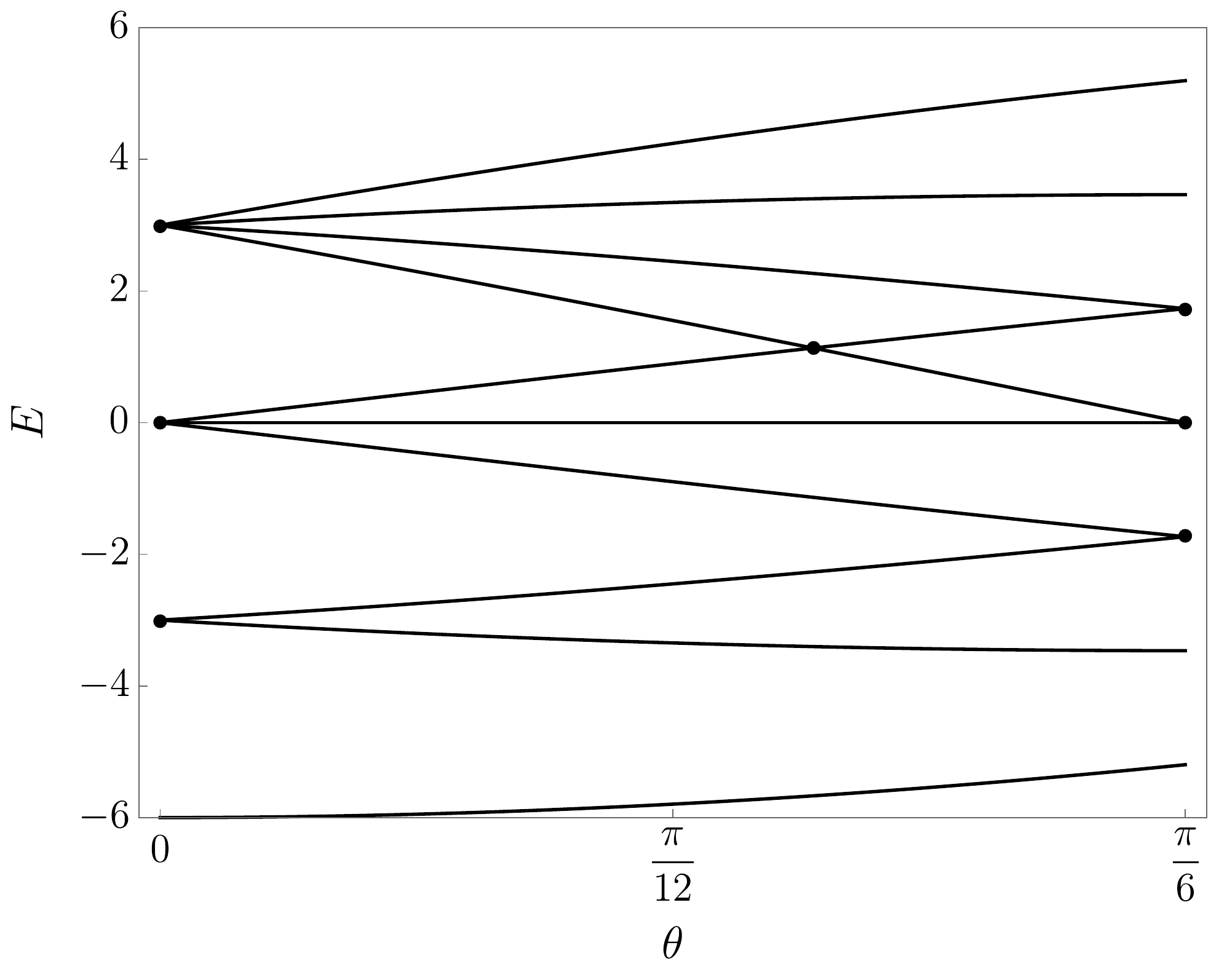}}
	\caption{
	(a) Free spectrum for a chain of length L=3, the only resonance points for such system are at $\theta=0$ and $\theta=\frac{\pi}{6}$.
	 (b) Free spectrum for a chain of length L=4, by increasing the length of the system a new resonant point appears at $\theta=\arctan{\frac{\sqrt{3}}{5}}$.
	 }
\label{fig::ChainLength34}
\end{figure}

One can show that if condition (\ref{eq::conditionExistence}) does not hold, then it is generally not possible to find a solution of the perturbative expansion for $\psi_p^{(k)}$ such that condition (\ref{eq::BrillouinWigner1Identity}) is satisfied and hence a local solution for the formal expansion of the zero mode does not exist. This fact is a generalization of what happens for the case $\theta=0$ that we analysed in the previous section, and it has to do with the fact that the identity operator $I_1$ contains a sum of spins as in (\ref{eq::identity}). For more details we refer to Appendix \ref{appendix:ProofProperty5}.

Condition (\ref{eq::conditionExistence}) is generally not satisfied when total domain wall angle is not conserved, making it also a sufficient condition for the conservation of the total domain wall angle. 

This raises now questions on the order at which the expansion breaks down when we are at resonant points. As we saw, when finding the zero mode expansion, we effectively consider chains of growing lengths. In this sense new resonance points appear as we consider chains of increasing length. Consider for example the spectrum of chains of length $L=3$ and $L=4$ in Figure \ref{fig::ChainLength34}, where we see that new resonant points appear. This means that an expansion of the zero mode, if it exists, can exist only up to an order that is equal to the length of the chain where the resonant point first appears. This agrees with the perturbative analysis conducted in \cite{OurPaper}.

We can sum up this discussion with the following 
\begin{property}
\label{property::NonExistence}
Suppose that $\theta$ is at a resonant point which does not conserve the total domain wall angle. Then the formal expansion for the zero mode can exist only up to an order compatible with the length of the chain at which the resonant point first appears.
\end{property}

This shines some light into the behaviour of the zero mode expansion when total domain wall angle is not conserved, but still fails to address the problem of the existence of zero modes at those values of $\theta$ where the total domain wall angle is conserved (including all non-resonant points). Unfortunately we are not able to provide a complete proof of the existence of zero modes in these cases. However an extensive analysis conducted through symbolic calculation with Mathematica shows that, when condition (\ref{eq::conditionExistence}) is satisfied, a formal expansion of the zero mode at each order $k$ exists, and can be constructed systematically. A procedure for this construction, which we conjecture to be an algorithm, is presented in section~\ref{sec::AlgorithmicSolution}. The solutions for the zero mode present interesting general properties that are extend readily to other spin chain models. This will be the subject of the next section.

\section{General form of the solution}
\label{sec::GeneralFormOfTheSolution}

We will now present a method that allows to construct a general solution for the recursive problem. We stress that our approach can be used to consider the solution of similar problems in other spin chain models as well, which we explored to some extent and we will discuss later in the section. This hints at the existence of some general principle that allows the machinery to work, although some of the details are still mysterious to us. 

As we already pointed out, the problem of finding a zero mode at some order $k$ is essentially a problem of finding $\psi_p^{(k)}$. For clarity we repeat that once we find a $\psi_p^{(k)}$ such that
\begin{equation}
    \mathcal{P}_0\mathcal{V}(\psi_p^{(k)}+\psi_q^{(k)})\otimes I_1=0
\end{equation}
we can readily find $\psi_q^{(k+1)}$, by inverting $\mathcal{H}_0$
\begin{equation}
    \psi_q^{(k+1)}=-\frac{\mathcal{Q}_0}{\mathcal{H}_0}\mathcal{V}(\psi_p^{(k)}+\psi_q^{(k)})\otimes I_1
\end{equation}
and this inversion is straightforward when everything is written in the basis that we chose, as $\mathcal{H}_0$ is in diagonal form.

As we saw the problem in finding $\psi_p^{(k)}$ is implicitly connected to the locality of the operator $\mathcal{V}\psi_q^{(k)}\otimes I_1$, as described by condition (\ref{eq::conditionExistence}). If condition (\ref{eq::conditionExistence}) is satisfied, then we find that, to the extent that it was possible to check, the solution of the problem can be written in the form
\begin{equation}
\label{eq::solutionChip}
\psi_p^{(k)}=\sum_{l_1+l_2+\ldots+l_{k-1}=k}\Gamma_{l_1,l_2,\ldots,l_{k-1}}\mathcal{P}_0\mathcal{V}\mathcal{S}^{l_1}\mathcal{V}\mathcal{S}^{l_2}\cdots\mathcal{S}^{l_{k-1}}\mathcal{V}\psi_p^{(0)}
\end{equation}
with $\Gamma_{l_1,l_2,\ldots,l_{k-1}}\in \mathbb{Q}$ and
\begin{equation}
    \mathcal{S}^l=
    \begin{cases}
    &\ \ \mathcal{P}_0\qquad l=0\\
    &\left(\frac{\mathcal{Q}_0}{\mathcal{H}_0}\right)^l\quad l\neq 0
    \end{cases}
\end{equation}
In the case of the $N=3$ parafermionic clock model we have, for example:
\begin{flalign*}
        \psi_p^{(0)}&=\sigma_1\ ,\hspace{14cm}\\
        \psi_p^{(1)}&=0\ ,\\
        \psi_p^{(2)}&=-\frac{1}{2}\mathcal{P}_0\mathcal{V}\left(\frac{\mathcal{Q}_0}{\mathcal{H}_0}\right)^2\mathcal{V}\psi_p^{(0)}\ ,\\
        \psi_p^{(3)}&=\frac{1}{2}\mathcal{P}_0\mathcal{V}\left(\frac{\mathcal{Q}_0}{\mathcal{H}_0}\right)^2\mathcal{V}\left(\frac{\mathcal{Q}_0}{\mathcal{H}_0}\right)\mathcal{V}\psi_p^{(0)}+\frac{1}{2}\mathcal{P}_0\mathcal{V}\left(\frac{\mathcal{Q}_0}{\mathcal{H}_0}\right)\mathcal{V}\left(\frac{\mathcal{Q}_0}{\mathcal{H}_0}\right)^2\mathcal{V}\psi_p^{(0)}\ ,
\end{flalign*}
\begin{equation*}
    \begin{split}
        &\psi_p^{(4)}=\frac{3}{8}\mathcal{P}_0\mathcal{V}\left(\frac{\mathcal{Q}_0}{\mathcal{H}_0}\right)^2\mathcal{V}\mathcal{P}_0\mathcal{V}\left(\frac{\mathcal{Q}_0}{\mathcal{H}_0}\right)^2\mathcal{V}\psi_p^{(0)}-\frac{3}{4}\mathcal{P}_0\mathcal{V}\left(\frac{\mathcal{Q}_0}{\mathcal{H}_0}\right)^2\mathcal{V}\left(\frac{\mathcal{Q}_0}{\mathcal{H}_0}\right)\mathcal{V}\left(\frac{\mathcal{Q}_0}{\mathcal{H}_0}\right)\mathcal{V}\psi_p^{(0)}+\\
        &\phantom{\psi_p^{(4)}}-\frac{1}{2}\mathcal{P}_0\mathcal{V}\left(\frac{\mathcal{Q}_0}{\mathcal{H}_0}\right)\mathcal{V}\left(\frac{\mathcal{Q}_0}{\mathcal{H}_0}\right)^2\mathcal{V}\left(\frac{\mathcal{Q}_0}{\mathcal{H}_0}\right)\mathcal{V}\psi_p^{(0)}-\frac{1}{4}\mathcal{P}_0\mathcal{V}\left(\frac{\mathcal{Q}_0}{\mathcal{H}_0}\right)\mathcal{V}\left(\frac{\mathcal{Q}_0}{\mathcal{H}_0}\right)\mathcal{V}\left(\frac{\mathcal{Q}_0}{\mathcal{H}_0}\right)^2\mathcal{V}\psi_p^{(0)}+\\
        &\phantom{\psi_p^{(4)}}+\frac{1}{4}\mathcal{P}_0\mathcal{V}\mathcal{P}_0\mathcal{V}\left(\frac{\mathcal{Q}_0}{\mathcal{H}_0}\right)\mathcal{V}\left(\frac{\mathcal{Q}_0}{\mathcal{H}_0}\right)^3\mathcal{V}\psi_p^{(0)}+\frac{3}{4}\mathcal{P}_0\mathcal{V}\mathcal{P}_0\mathcal{V}\left(\frac{\mathcal{Q}_0}{\mathcal{H}_0}\right)^2\mathcal{V}\left(\frac{\mathcal{Q}_0}{\mathcal{H}_0}\right)^2\mathcal{V}\psi_p^{(0)}\ .
    \end{split}
\end{equation*}
For the values taken by the $\Gamma$s in the next order that we could check we refer to Appendix \ref{appendix::GammasForPsip5}. Note that the coefficients do not depend on $\theta$ and $\phi$ and that all the terms in each sum act trivially on the last site of the chain, as imposed by Property \ref{property::conditionExistence}. This gives another hint of the fact that when $\theta$ is such that the total domain-wall angle is not conserved, the zero mode operator cannot be constructed, as the presence of $\frac{\mathcal{Q}_0}{\mathcal{H}_0}$ will induce divergences at resonant points that cannot be removed. In particular this accounts for the divergences witnessed in the expansion of the zero mode in \cite{Fendley2012,Kemp_2017,OurPaper}. 

In Appendix \ref{appendix:ProofChip} we present a proof for the solutions of $\psi_p^{(2)}$. The solutions for the next orders were checked by symbolic computation using Mathematica. We were able to explicitly check our ansatz up to $5^{\mathrm{th}}$ order in the perturbative expansion\footnote{The solution for $\psi_{p}^{(5)}$ is given in Appendix~\ref{appendix::GammasForPsip5}}, even though we know that a solution for $\psi_p^{(6)}$ exists. (We were able to construct this solution using the methods in section~\ref{sec::AlgorithmicSolution}, but not able to check that it takes the form conjectured in this section.) Note that already to the third order we have to deal with a Hilbert space of dimension $3^6$ and that the existence of these types of solutions is therefore related to the exact cancellation of a very large number of terms, making it hard to believe that the above structure is simply the result of chance. The results from \cite{OurPaper} and \cite{Else2017} corroborates this idea. This becomes even more apparent when we try to apply the same methodology to other spin models.  

As a final remark we note that this ansatz is quite reminiscent of the formulas obtained for effective Hamiltonians in the framework of degenerate perturbation theory \cite{Soliverez1969,Klein1974}.

\subsection{Iterative method in other models}

In order to further understand the structure of the solution we can also apply our method to spin models where the existence of a zero mode is already known, like the XYZ model \cite{XYZ} or the models considered in \cite{Kemp_2017}. In the cases we studied, we found that the known solutions for the zero modes follow our description. Interestingly the existence of these formal expressions for the zero modes hold even when there is no underlying symmetry, as for example in the case of the Hamiltonian\footnote{This model was considered in several other works, see e.g.~ \cite{Ovchinnikov2003,Dutta2015}.}
\begin{equation}
    H'=-\sum_{i=1}^{L-1} \sigma_{i}^z\sigma_{i+1}^z+f\sum_{i}^L\sigma_i^{x}+\sigma_i^{z}\ .
\end{equation}
with $f$ the perturbing parameter. This Hamiltonian does not commute with the fermion parity (or with any other symmetry, as far as we are aware) and is not integrable. Nonetheless our construction can still be carried out starting with $\psi_p^{(0)}=\sigma_1^z$ (the free model still possesses a local zero mode on the left edge). The resulting zero mode is in general not normalizable, which means that we do not expect it to survive in the thermodynamic limit, but surprisingly a formal expression for it can still be written out for any order we could check (up to 8th order). It might be interesting to consider how this is related to the slow thermal relaxation of local operators treated in \cite{Hyungwon2015}.

By considering different models we also see that in general the coefficients $\Gamma$s are model dependent. For example if we consider the Hamiltonian
\begin{equation}
     H'=-\sum_{i=1}^{L-1} \sigma_{i}^z\sigma_{i+1}^z+f\sum_{i}\sigma_{i+1}^{x}\sigma_i^x+\sigma_{i+1}^{y}\sigma_i^y=H'_0+fV'
\end{equation}
the first orders of the projections of the zero mode, $\psi'_p\in Null(\mathcal{H'}_0)$, constructed with our method, are given by
\begin{equation*}
\small
    \begin{split}
        &{\psi'}_p^{(0)}=\sigma_1^z\ ,\\
        &{\psi'}_p^{(1)}=0\ ,\\
        &{\psi'}_p^{(2)}=-\frac{1}{2}\mathcal{P'}_0\mathcal{V'}\left(\frac{\mathcal{Q'}_0}{\mathcal{H'}_0}\right)^2\mathcal{V'}{\psi'}_p^{(0)}\ ,\\
        &{\psi'}_p^{(3)}=-\frac{2}{3}\mathcal{P'}_0\mathcal{V'}\left(\frac{\mathcal{Q'}_0}{\mathcal{H'}_0}\right)^2\mathcal{V'}\left(\frac{\mathcal{Q'}_0}{\mathcal{H'}_0}\right)\mathcal{V'}\psi_p^{(0)}-\frac{1}{3}\mathcal{P'}_0\mathcal{V'}\left(\frac{\mathcal{Q'}_0}{\mathcal{H'}_0}\right)\mathcal{V'}\left(\frac{\mathcal{Q'}_0}{\mathcal{H'}_0}\right)^2\mathcal{V'}{\psi'}_p^{(0)}\ .
    \end{split}
\end{equation*}

We believe that the existence of these expressions has to do with the chiral symmetry $\mathcal{TK}$ introduced in Section \ref{sec::Symmetry}, however we leave this for future work.

 Finally we tried to apply the same method also to non-hermitian models and the same general structure holds. Notably we investigated the case of free parafermions \cite{Fendley2014}, which can be described through the Hamiltonian $H'=H'_0+fV'$, with
\begin{equation}
    \begin{split}
        &H'=-\sum_{i=1}^{L-1}\sigma_{i+1}^\dagger\sigma_i-f\sum_{i=1}^L\tau_i\ .
    \end{split}
\end{equation}
This is the same as the spin clock model (\ref{eq::InitialHamiltonian}) when $\theta=0$ and $\phi=0$, except for the omission of the hermitian conjugate terms. In this case, to each order $k$, we find
\begin{equation}
   {\psi'}^{(k)}_q=\left(-\frac{\mathcal{Q}_0}{\mathcal{H'}_0}\mathcal{V'}\right)^k\psi_p^{(0)}\qquad{\psi'}_p^{(k)}=0\ .
\end{equation}
These are formally the same expressions that one obtains in the case of the transverse Ising model ($N=2$) and this provides further evidence that these types of models, rather than (\ref{eq::InitialHamiltonian}), constitute a closer generalization, even if not hermitian, of the transverse Ising model \cite{Fendley2014}.

\section{The problem of normalization}
\label{sec::Normalization}

In the last section we showed that there are cases where we can write down a formal expression for the zero mode. However the fact that we can construct these operators does not generally mean that in the limit $L\to\infty$ the zero modes exists, as the perturbation series could fail to converge. In this section we will therefore address the problem of normalization. As outlined in Section \ref{section::parafermionicZeroMode}, we require our zero mode to satisfy the conditions
\begin{equation}
\label{eq::normalizationCube}
    \psi^3=I_L+O\left(f^{-L}\right)
\end{equation}
and
\begin{equation}
\label{eq::normalizationDagger}
    \psi^\dagger\psi=I_L+O\left(f^{-L}\right)\ .
\end{equation}
In this section we will show that these two conditions are equivalent and that once we find a formal solution for the zero mode they can always be fulfilled. Let's therefore consider these cases in full generality.

\subsection{Expansions of $\psi^{3}$}
Consider $\psi^3$ first. Since $\psi$ admits an expansion in $f$, so does $\psi^3$:
\begin{equation}
    \psi^3=I_L+f(\psi^3)^{(1)}+f^2(\psi^3)^{(2)}+\ldots
\end{equation}
and we have 
\begin{equation}
    (\psi^3)^{(k)}=\sum_{\substack{k_1+k_2+k_3=k}}\psi^{(k_1)}\psi^{(k_2)}\psi^{(k_3)}\ .
\end{equation}
Since every $\psi^{(k)}$ satisfies the equation $[H_0,\psi^{(k)}]=-[V,\psi^{(k-1)}]$ so does $(\psi^3)^{(k)}$, in fact 
\begin{equation*}
    \begin{split}
   \left[H_0,(\psi^3)^{(k)}\right]=\sum_{\substack{k_1+k_2+k_3=k}}\left[H_0,\psi^{(k_1)}\right]\psi^{(k_2)}\psi^{(k_3)}+\psi^{(k_1)}\left[H_0,\psi^{(k_2)}\right]\psi^{(k_3)}+\psi^{(k_1)}\psi^{(k_2)}\left[H_0,\psi^{(k_3)}\right]
\end{split}
\end{equation*}
using that $\left[H_0,\psi^{(l)}\right]=-\left[V,\psi^{(l-1)}\right]$ for all $l\leq k$ we get
\begin{equation*}
    \begin{split}
        \sum_{\substack{k_1+k_2+k_3=k}}-\left[V,\psi^{(k_1-1)}\right]\psi^{(k_2)}\psi^{(k_3)}-\psi^{(k_1)}\left[V,\psi^{(k_2-1)}\right]\psi^{(k_3)}-\psi^{(k_1)}\psi^{(k_2)}\left[V,\psi^{(k_3-1)}\right]\ .
    \end{split}
\end{equation*}
Since $\psi^{(-1)}=0$, we see that 
\begin{equation}
    \left[H_0,(\psi^3)^{(k)}\right]=-\left[V,(\psi^3)^{(k-1)}\right]
\end{equation}
or, written in the super-operator formalism
\begin{equation}
\label{eq::etaPerturbation}
    \mathcal{H}_0(\psi^3)^{(k)}\otimes I_1=-\mathcal{V}(\psi^3)^{(k-1)}\otimes I_1\ .
\end{equation}

In line with what we have done in the previous sections we can project this equation down to $Null(\mathcal{H}_0)$ and its orthogonal space. Equation (\ref{eq::etaPerturbation}) is therefore equivalent to 
\begin{equation}
\label{eq::JordanWignerCube}
\begin{split}
    &\mathcal{H}_0(\psi^3)^{(k)}_q\otimes I_1=-\mathcal{Q}_{0}\mathcal{V}\left((\psi^3)^{(k-1)}_q+(\psi^3)^{(k-1)}_p\right)\otimes I_1\\
    &\mathcal{P}_0\mathcal{V}(\psi^3)^{(k)}_p\otimes I_1=-\mathcal{P}_0\mathcal{V}(\psi^3)^{(k)}_q\otimes I_1
    \end{split}
\end{equation}
with $(\psi^3)^{(k)}_q=\mathcal{Q}_0(\psi^3)^{(k)}$ and $(\psi^3)^{(k)}_p=\mathcal{P}_0(\psi^3)^{(k)}$. We can now see that the following property holds
\begin{property}
\label{property::normalizationCube}
If a solution for the zero mode exists up to some order $k$ then
\begin{equation}
    (\psi^3)^{(j)}_q=0\qquad(\psi^3)^{(j)}_p=\lambda_jI_{j+1} \qquad\forall j=1,2\ldots,k
\end{equation}
with constants $\lambda_j\in\mathbb{C}$.
\end{property}
\noindent This property can be easily proved by induction. It is true for $j=0$, as $$(\psi^3)^{(0)}=\sigma_1^3=I_1\ .$$ 
Since $$ \mathcal{Q}(\psi^3)^{(0)}=(\psi^3)^{(0)}$$ and since the super-operator $\mathcal{Q}$ commutes with $\mathcal{H}_0$ and $\mathcal{V}$, from (\ref{eq::JordanWignerCube}), we can assume that
\begin{equation}
    \label{eq::proofNormalizationTemp}
     \mathcal{Q}(\psi^3)^{(j)}=(\psi^3)^{(j)}\qquad \forall j=1,2,\ldots,k\ .
\end{equation}
Suppose therefore that the Property \ref{property::normalizationCube} is true for $j-1$, that is 
\begin{equation}
    (\psi^3)^{(j-1)}_q=0\qquad (\psi^3)^{(j-1)}_p=\lambda_{j-1}I_j\ .
\end{equation}
Since $\mathcal{V}I_j=0$ we have
\begin{equation*}
    \mathcal{H}_0(\psi^3)^{(j)}_q\otimes I_1=-\mathcal{Q}_{0}\mathcal{V}(\psi^3)^{(j-1)}_p\otimes I_1=0\ ,
\end{equation*}
which in turn means 
$$(\psi^3)^{(j)}_q=0\ ,$$ 
because by hypothesis $(\psi^3)^{(j)}_q\notin Null(\mathcal{H}_0)$. Hence we are left with
\begin{equation*}
    \mathcal{P}_0\mathcal{V}(\psi^3)^{(j)}_p\otimes I_1=-\mathcal{P}_0\mathcal{V}(\psi^3)^{(j)}_q\otimes I_1=0\ .
\end{equation*}
We already encountered this equation in Section \ref{sec::RestrictionOnTheSolution} and we know that its solutions are the linear combinations of the operators
\begin{equation*}
    |0\rangle|0\rangle\otimes I_{j}\qquad|1\rangle|1\rangle\otimes I_{j}\qquad|2\rangle|2\rangle\otimes I_{j}\ .
\end{equation*}
Since we are supposing that a solution exists and because of (\ref{eq::proofNormalizationTemp}), we have that the unique solution is given by
\begin{equation}
    (\psi^3)^{(j)}_p=\lambda_jI_j\ .
\end{equation}
for some constant $\lambda_j\in \mathbb{C}$, that will generally depend on $\theta$.

\subsection{Expansion of $\psi^\dagger\psi$}

We can now consider what happens to $\psi^\dagger\psi$. Also in this case, since $\psi$ admits an expansion in $f$, so does $\psi^\dagger\psi$
\begin{equation}
    \psi^\dagger\psi=I_L+f(\psi^\dagger\psi)^{(1)}+f^2(\psi^\dagger\psi)^{(2)}+\ldots
\end{equation}
and we have
\begin{equation}
    (\psi^\dagger\psi)^{(k)}=\sum_{\substack{k_1+k_2=k}}{\psi^{(k_1)}}^\dagger\psi^{(k_2)}\ .
\end{equation}
Now the discussion goes exactly like the previous one for $\psi^3$ and we have that $(\psi^\dagger\psi)^{(k)}$ satisfies the equations
\begin{equation}
    \label{eq::JordanWignerDagger}
    \begin{split}
    &\mathcal{H}_0(\psi^\dagger\psi)^{(k)}_q=-\mathcal{Q}_{0}\mathcal{V}\left((\psi^\dagger\psi)^{(k-1)}_q+(\psi^\dagger\psi)^{(k-1)}_p\right)\otimes I_1\\
    &\mathcal{P}_0\mathcal{V}(\psi^\dagger\psi)^{(k)}_p\otimes I_1=-\mathcal{P}_0\mathcal{V}(\psi^\dagger\psi)^{(k)}_q\otimes I_1\ .
    \end{split}
\end{equation}
In the same way as before, the following property can be proved
\begin{property}
\label{property::normalizationDagger}
If a solution for the zero mode exists up to some order $k$ then
\begin{equation}
    (\psi^\dagger\psi)^{(j)}_q=0\qquad(\psi^\dagger\psi)^{(j)}_p=\lambda'_jI_L \qquad\forall j=1,2\ldots,k
\end{equation}
with constants $\lambda'_j\in\mathbb{R}$.
\end{property}

\subsection{Expansions of $\psi^2$ and $\psi^\dagger$}

We will now consider the expansions of $\psi^2$ and $\psi^\dagger$~\footnote{Note that in the super-operator basis (\ref{eq::basis}) $\psi^\dagger$ is given by
\begin{equation*}
    \psi^\dagger=\mathcal{TK}\psi
\end{equation*}
} and we will see how we can choose the normalization by considering the relation between them. To understand the point suppose that
\begin{equation}
    \psi^2=\psi^\dagger+O(f^L)\ .
\end{equation}
then this would imply that 
\begin{equation}
    (\psi^3)^{(j)}=(\psi^\dagger\psi)^{(j)}\qquad \forall\ j=1,2,\ldots,L-1
\end{equation}
and therefore:
\begin{equation}
    \lambda_j=\lambda'_j\qquad\forall\ j=1,2,\ldots,L-1\ .
\end{equation}
If this condition is satisfied then we can always make sure that (\ref{eq::normalizationCube}, \ref{eq::normalizationDagger}) are satisfied by renormalizing $\psi$. This therefore raises the question: can we make sure that at each order $j$
\begin{equation*}
    (\psi^2)^{(j)}=(\psi^{(j)})^\dagger\ ,
\end{equation*}
and if not, can we use the freedom in choosing a solution to make sure that this becomes true? We can start to answer this question by establishing the following 
\begin{property}
\label{property::psiSquare}
Suppose that
\begin{equation}
\label{eq::psidaggerpsi2order}
    (\psi^2)^{(j)}=(\psi^{(j)})^\dagger
\end{equation}
with $j=1,2,\ldots,k-1$. Then we have
\begin{equation}
    (\psi^2)^{(k)}-(\psi^\dagger)^{(k)}=\lambda''_k\sigma_1^\dagger\otimes I_{L-1}
\end{equation}
with $\lambda''_k\in\mathbb{C}$.
\end{property}
\noindent The proof of this property goes exactly in the same way as the proofs for  Property \ref{property::normalizationCube} and Property \ref{property::normalizationDagger}. Although note that for it to be true we need to assume that we are able, somehow, to make sure that (\ref{eq::psidaggerpsi2order}) holds at each order $j<k$. The only difference is that 
\begin{equation*}
    \mathcal{Q}((\psi^2)^{(0)}-(\psi^\dagger)^{(0)})=\omega^2((\psi^2)^{(0)}-(\psi^\dagger)^{(0)})
\end{equation*}
which means that a solution of
\begin{equation*}
    \mathcal{P}_0\mathcal{V}((\psi^2)^{(k)}-(\psi^\dagger)^{(k)})\otimes I_1=0\ .
\end{equation*}
is given by $\lambda''_k\sigma_1^2\otimes I_{k}$. We are finally in the position to consider how to fix the normalization.

\subsubsection{Choice of the normalization}

As we already pointed out, whenever we find a solution for the expansion of the zero mode, it is not unique. In fact, given a solution of order $k$, we can always add to it a solution of the equation
\begin{equation}
    \mathcal{P}_0\mathcal{V}\xi_p^{(k)}\otimes I_1=0\ ,
\end{equation}
with $\xi_p^{(k)}\in Null(\mathcal{H}_0)$ and we would still get a zero mode, in fact
\begin{equation*}
    \mathcal{P}_0\mathcal{V}(\psi_p^{(k)}+\xi_p^{(k)})\otimes I_1=\mathcal{P}_0\mathcal{V}\psi_p^{(k)}\otimes I_1=-\mathcal{P}_0\mathcal{V}\psi_q^{(k)}\otimes I_1\ .
\end{equation*}
By now we should be acquainted with the fact that if we are looking for solutions such that $\mathcal{Q}\psi=\omega\psi$, the only possible choice for $\xi_p^{(k)}$ is given by
\begin{equation}
    \xi_p^{(k)}=\xi_k\sigma_1\otimes I_{k}
\end{equation}
with $\xi_k\in\mathbb{C}$. 

This freedom can be used to fix the normalization and in particular we can use it to enforce the condition
\begin{equation}
     (\psi^2)^{(j)}=(\psi^\dagger)^{(j)} .
\end{equation}
Doing this at every order will guarantee that
\begin{equation}
    \psi^2=\psi^\dagger+O(f^L)\ .
\end{equation}
First of all note that 
\begin{equation}
    (\psi^{2})^{(k)}=\sum_{k_1+k_2=k}\psi^{(k_1)}\psi^{(k_2)}\ .
\end{equation}
It is therefore not difficult to prove that if we use our freedom to add operators of the form $\xi_p^{(k)}$ to $\psi^{(k)}$
\begin{equation}
    \psi^{(k)}\to\psi^{(k)}+\xi_p^{(k)}
\end{equation}
$(\psi^2)^{(k)}-(\psi^\dagger)^{(k)}$ will transform as
\begin{equation}
    (\psi^2)^{(k)}-(\psi^\dagger)^{(k)}\to (\psi^2)^{(k)}-(\psi^\dagger)^{(k)}+(2\xi_k-\xi_k^*)\sigma_1^\dagger
\end{equation}
By using induction and Property \ref{property::psiSquare}, we can therefore see that if at every order $j\leq L-1$ we set
\begin{equation}
    \xi_j=-\text{Re}(\lambda''_j)-\frac{i}{3}\text{Im}(\lambda''_j)\ ,
\end{equation}
then 
\begin{equation}
    \psi^2=\psi^\dagger+O\left(f^L\right)\ .
\end{equation}
As we already noted, because of Property \ref{property::normalizationCube} and Property \ref{property::normalizationDagger} this condition implies
\begin{equation}
    (\psi^3)^{(j)}=\lambda_jI_L\qquad(\psi^\dagger\psi)^{(j)}=\lambda_jI_L\qquad \forall j=1,2,\ldots,k
\end{equation}
and $\lambda_j\in\mathbb{R}$ as in Property \ref{property::normalizationDagger}. Therefore, up to a common normalization factor, equations (\ref{eq::normalizationCube}) and (\ref{eq::normalizationDagger}) are true. Note again that in general $\lambda_j$ will depend on $\theta$ and $\phi$.

\begin{figure}[t!]
\centering
 \subfigure[]{\includegraphics[scale=0.2993]{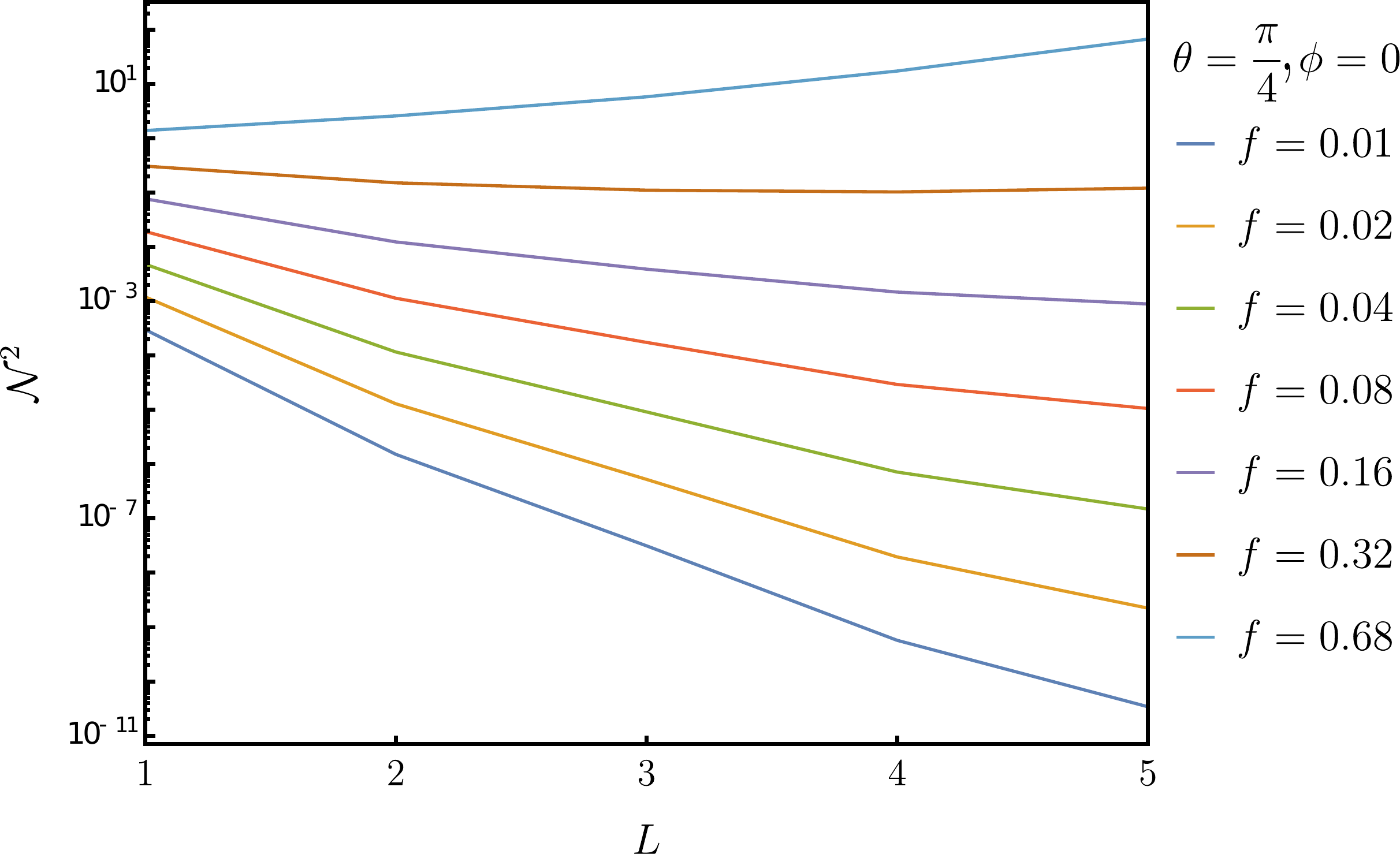}
 \label{fig::NormalizationPi4}}
\subfigure[]{\includegraphics[scale=0.2993]{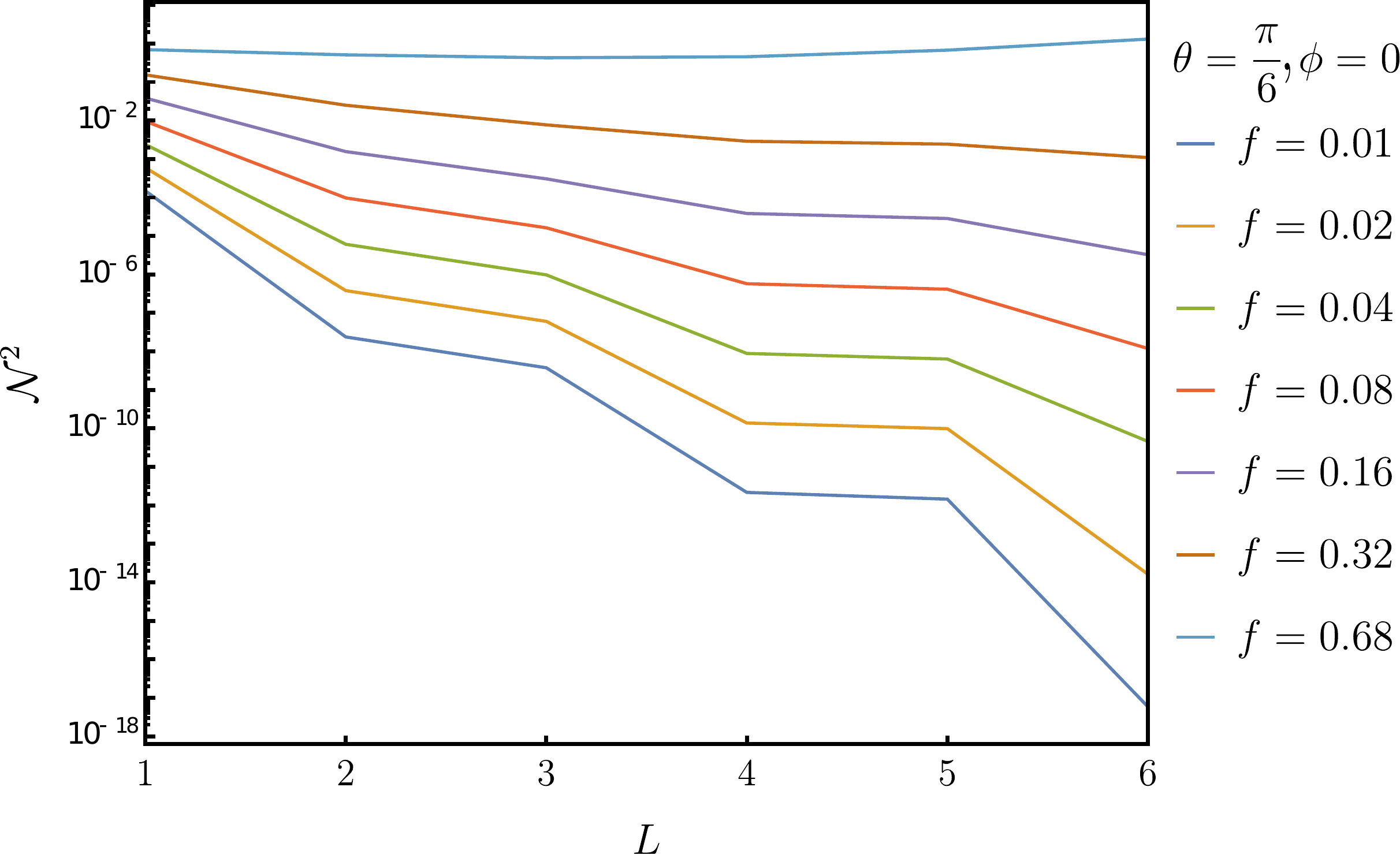}\label{fig::NormalizationPi6}}
    \caption{In (a) we show the value of $\mathcal{N}^2$ in equation~(\ref{eq:normdeviation}), which signals the deviation from $1$ of the norm of the truncated expansion, for different chain lengths $L$ and $\theta=\frac{\pi}{4}$, $\phi=0$. In (b) we show $\mathcal{N}^2$  for different chain lengths $L$ and $\theta=\frac{\pi}{6}$, $\phi=0$.}
    \label{fig::Normalization}
\end{figure}
\section{Convergence of the formal series}
\label{sec::convergenceSeries}
Even though the method outlined above seems to work at every order, divergences may still arise, as the coefficients of the truncated series constituting the zero mode for a chain of length $L$ could grow faster than $f^{-L}$ . In this sense the above expansions for the zero mode have to be to considered as formal expressions that we can write whenever $\theta$ is such that the total domain wall angle is conserved. 
\\The next problem we need to address is therefore about the radius of convergence of these formal series in $f$.

To this end we first need to define what is the error that we make when we truncate the expressions for the zero modes.
Hence we define
\begin{equation}
\label{eq:normdeviation}
    \mathcal{N}^2=\left|\frac{1}{3^L}\text{Tr}\left(\psi^\dagger\psi\right)-1\right|\ ,
\end{equation}
where the normalization factor $1/3^L$ is such that $I_L$ has norm $1$ and $\psi$ is obtained by truncating the expansion at order $L$
\begin{equation}
    \psi=\psi^{(0)}+f\psi^{(1)}+\ldots+f^{L-1}\psi^{(L-1)}+f^L\psi^{(L)}\ .
\end{equation}
$\mathcal{N}^2$ is therefore simply the Frobenius norm of $\psi$. 

Even though we can always satisfy
\begin{equation}
    \psi^\dagger\psi=I_L+O\left(f^L\right)
\end{equation}
the sum of all terms of order $f^L$ could still diverge for $L\to\infty$. The expansion converges if and only if $\mathcal{N}^2\to0$. In Figure \ref{fig::Normalization} we show the plots of $\mathcal{N}^2$ for $\theta=\frac{\pi}{4}$ and $\frac{\pi}{6}$, in both cases we chose $\phi=0$. From these graphs we see that there seems to be a finite radius of convergence, even though the information available is limited.

A related problem concerns the convergence of the commutator between the total hamiltonian $H$ and the truncated zero mode 

\begin{figure}[t]
\centering
 \subfigure[]{\includegraphics[scale=0.2993]{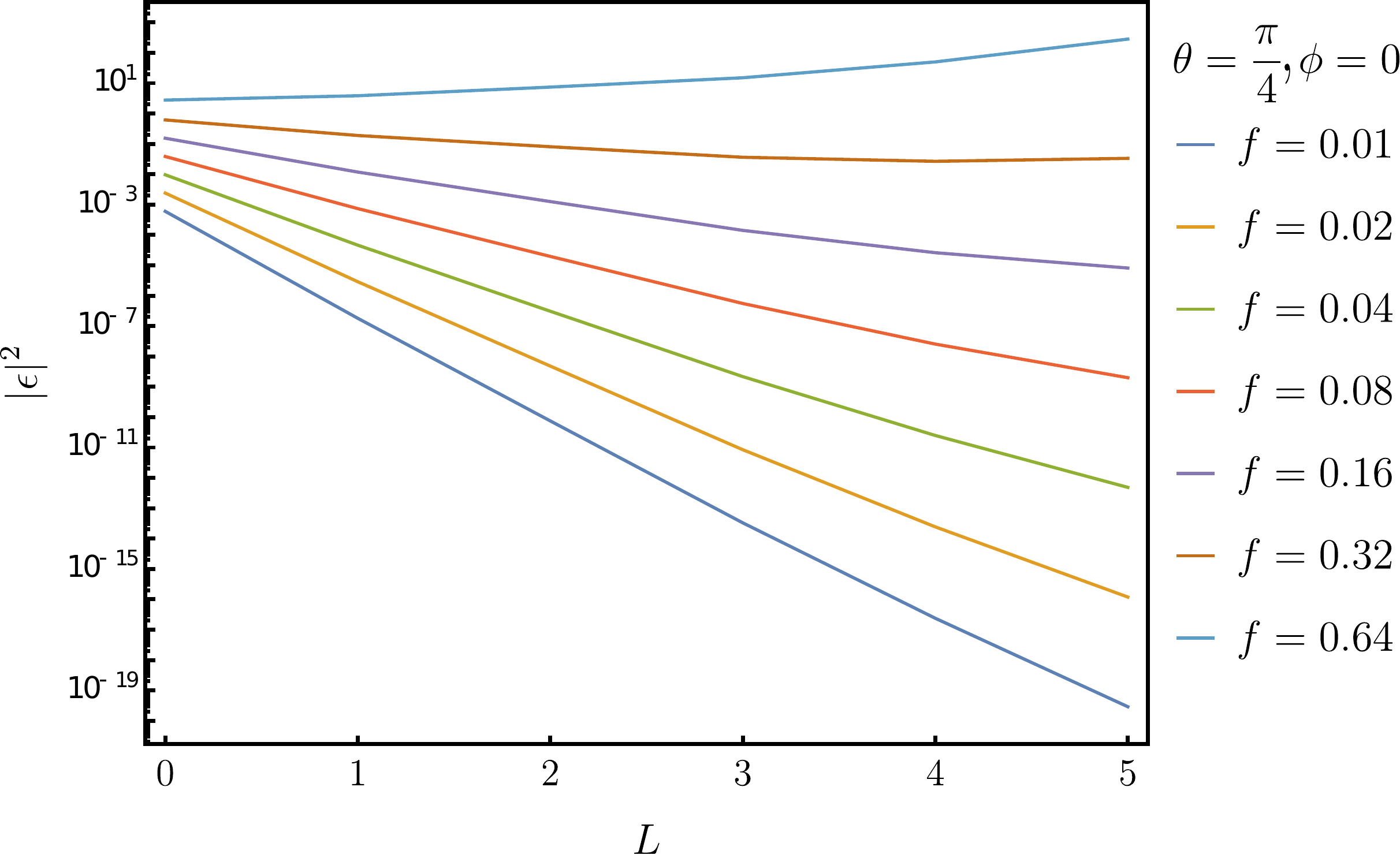}
 \label{fig:ErrorPi4}}
\subfigure[]{\includegraphics[scale=0.2993]{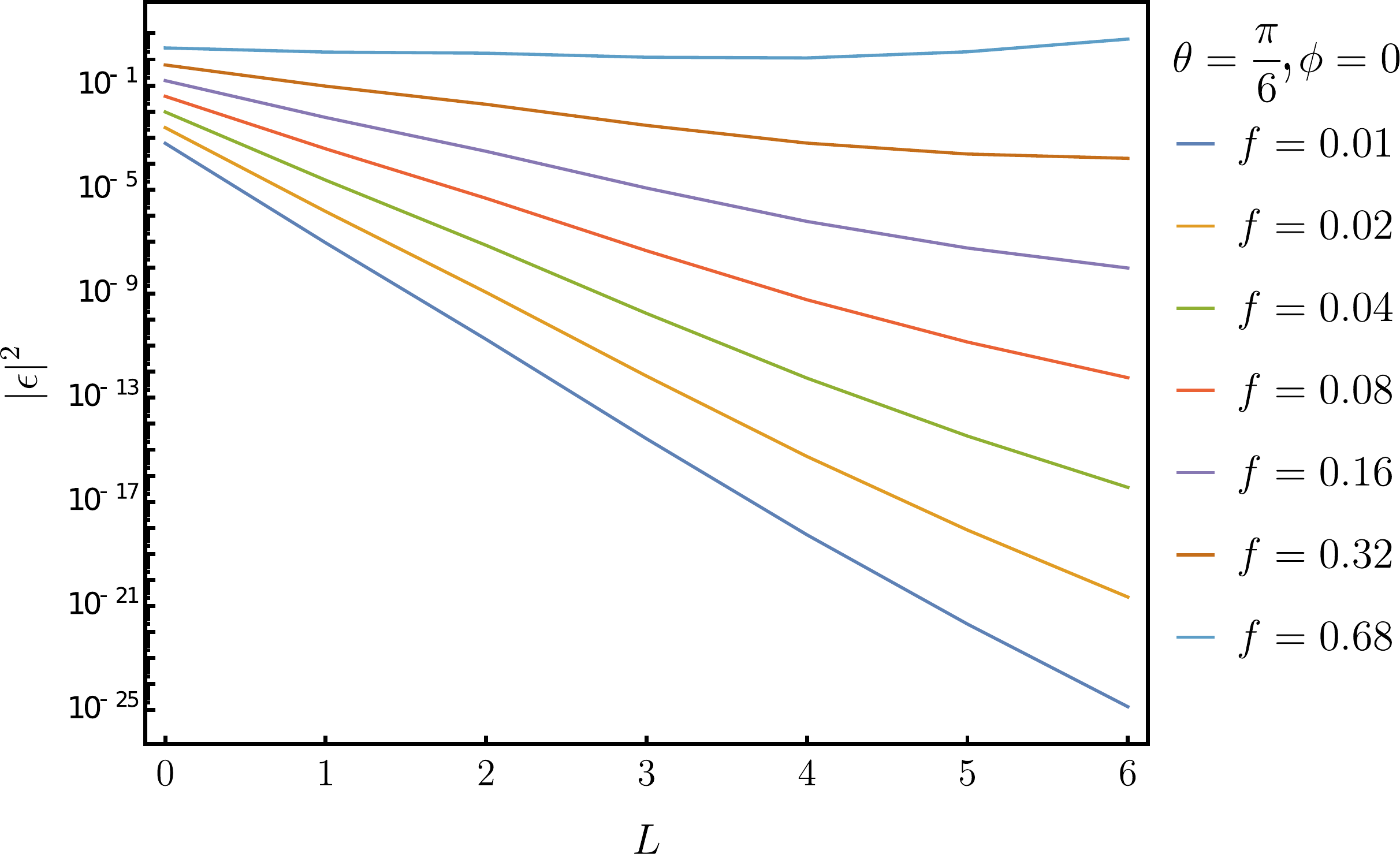}\label{fig::ErrorPi6}}
    \caption{In (a) we show the norm of $\epsilon$ in equation~(\ref{eq::epsilon}), for different chain lengths $L$ and $\theta=\frac{\pi}{4}$, $\phi=0$. In (b) we show the norm of $\epsilon$ for different chain lengths $L$ and $\theta=\frac{\pi}{6}$, $\phi=0$.}
    \label{fig::Error}
\end{figure}

In Figure \ref{fig::Error} we show plots for the norm\footnote{As before we consider the normalization constant to be $\frac{1}{3^L}$.} of $\epsilon$ in the cases of $\theta=\frac{\pi}{6}$, $\theta=\frac{\pi}{4}$ and $\phi=0$.
In this case we also have strong suggestion that a finite radius of convergence exists, but further analysis is needed in order to have a definitive answer on the subject. In particular, in order to give an estimate of the radius of convergence, we need more information about the constants $\Gamma_{l_1,l_2,\ldots,l_{k-1}}$ appearing in (\ref{eq::solutionChip}) and at the moment we lack a proper understanding of how these constants arise in the solution.
\begin{equation}
\label{eq::epsilon}
\epsilon=(\mathcal{H}_0+f\mathcal{V})\psi=f^{L+1}\mathcal{V}\psi^{(L)}\ .
\end{equation}


\section{Algorithmic solution for the formal series}
\label{sec::AlgorithmicSolution}

The general ansatz for the solution $\psi^{(k)}$ that we have provided in Section \ref{sec::GeneralFormOfTheSolution}, lends itself well to direct checking. Nevertheless the problem of finding the constants $\Gamma_{l_1,l_2,\ldots,l_{k-1}}$ of (\ref{eq::solutionChip}) is generally not an easy task, especially for large $k$. However, if we are not interested in the specific values taken by these constants, it is possible to design algorithms that would allow to find the $\psi_p^{(k)}$ at any given order (with limitations due to computational power). Strictly speaking we cannot prove that this algorithm works at all orders, as we need to assume our ansatz (\ref{eq::solutionChip}) to hold true in order for it to work. If this is the case we can employ some of the symmetries of the problem, and some of the remarkable cancellations of terms that have to take place in order for such solution to exist, to simplify the determination of $\psi_p^{(k)}$. As such, the recipes that we will present in this section have always worked. 

With this in mind consider a solution for $\psi_p^{(k)}$ as in (\ref{eq::solutionChip}). It can be seen by induction, using repeatedly property (\ref{eq::propertyEnergy}) and the local structure of $\mathcal{V}$, that
\begin{equation}
\label{eq::formSolutionAlgorithm}
    \psi_p^{(k)}=\psi_{p,k}+\psi_{p,k-1}\otimes I_1+\ldots+\psi_{p,1}\otimes I_{k-1}
\end{equation}
where $\psi_{p,l}$ are operators acting on chains of length $l$:
\begin{equation}
\label{eq::chipl}
\begin{split}
    \psi_{p,l}&=\sum_{j, i}(\psi_l)_{i_1,\mathbf{j}_{l-2},i_l}^{i_1,\mathbf{i}_{l-2},i_l}|i_1,\mathbf{i}_{l-2},i_l\rangle|i_1,\mathbf{j}_{l-2},i_l\rangle \quad\text{with}\quad E^{i_1,\mathbf{i}_{l-2},i_l}_{i_1,\mathbf{j}_{l-2},i_l}=0\ .
\end{split}
\end{equation}
With $\mathbf{i}_{l-2}$ and $\mathbf{j}_{l-2}$ not necessarily different from each other. The equation satisfied by the solution $\psi_p^{(k)}$ is given by
\begin{equation}
    \mathcal{P}_0(\mathcal{V}\psi^{(k)}_p\otimes I_1)=\beta_p^{(k)}
\end{equation}
with $\beta_p^{(k)}=-\mathcal{P}_0(\mathcal{V}\psi_q^{(k)})$ and
\begin{equation}
    \beta_p^{(k)}=\sum_{j, i}(\beta)_{i_1,\mathbf{j}_{k-1},i_{k+1}}^{i_1,\mathbf{i}_{k-1},i_{k+1}}|i_1,\mathbf{i}_{k-1},i_{k+1}\rangle|i_1,\mathbf{j}_{k-1},i_{k+1}\rangle \quad\text{with}\  E^{i_1,\mathbf{i}_{k-1},i_{k+1}}_{i_1,\mathbf{j}_{k-1},i_{k+1}}=0\ .
\end{equation}
The idea behind how to find a solution is now to determine one by one all the $\psi_{p,l}$ that constitute $\psi_p^{(k)}$, starting from $\psi_{p,k}$.
Doing this will reduce at each step the dimensionality of the problem. Suppose now that we are able to find $\psi_{p,k}$, then we can write
\begin{equation}
    \mathcal{P}_0(\mathcal{V}{\psi}^{(k,1)}_{p}\otimes I_2)={\beta}_p^{(k)}-\mathcal{P}_0(\mathcal{V}\psi_{p,k}\otimes I_1)
\end{equation}
with
\begin{equation}
    {\psi}_p^{(k,1)}=\psi_{p,k-1}+\psi_{p,k-2}\otimes I_1+\ldots+\psi_{p,1}\otimes I_{k-2}
\end{equation}
If a solution exists and follows (\ref{eq::formSolutionAlgorithm}), because of the locality of $\mathcal{V}$ and condition (\ref{eq::propertyEnergy}) then we must have
\begin{equation}
\mathcal{P}_0(\mathcal{V}\psi_p^{(k,1)}\otimes I_2)={\beta}_p^{(k,1)}\otimes I_1
\end{equation}
with ${\beta}_p^{(k,1)}\otimes I_1={\beta}_p^{(k)}-\mathcal{P}_0(\mathcal{V}\psi_{p,k}\otimes I_1)$ and
\begin{equation}
    {\beta}_p^{(k,1)}=\sum_{j, i}(\beta_1)_{i_1,\mathbf{j}_{k-2},i_{k}}^{i_1,\mathbf{i}_{k-2},i_{k}}|i_1,\mathbf{i}_{k-2},i_{k}\rangle|i_1,\mathbf{j}_{k-2},i_{k}\rangle \quad\text{with}\quad E^{i_1,\mathbf{i}_{k-2},i_{k}}_{i_1,\mathbf{j}_{k-2},i_{k}}=0\ .
\end{equation}
Therefore ${\beta}_p^{(k,1)}$ acts on a chain that is one site smaller than the one on which ${\beta}^{(k)}$ acts. This means that we can reduce the problem to
\begin{equation}
  \mathcal{P}_0(\mathcal{V}\psi_p^{(k,1)}\otimes I_1)={\beta}_p^{(k,1)}\ ,
\end{equation}
which is equivalent to solve the initial problem on a smaller chain and this is computationally very convenient as the Hilbert spaces involved in the process are of smaller dimensions. This process can be repeated now for $\psi_{p,k-1}$, $\psi_{p,k-2}$, etc...~until we exhaust all the terms.

We now need to describe how to determine the various $\psi_{p,l}$. We remark that the fact that it is always possible to go from $\beta_p^{(k,l)}$ living on a chain of length $k+1-l$ to a $\beta_p^{(k,l+1)}$ living on a chain of length $k-l$ is not trivial. It is a consequence of the existence of a solution of the form (\ref{eq::formSolutionAlgorithm}) and this implies the equality and the cancellation of a large number of terms in $\beta_p^{(k)}$ when we remove $\psi_{p,l}$ as described above. The presence of this structure can be traced back to the general form of the solution described in Section \ref{sec::GeneralFormOfTheSolution} and is part of the reason that makes us believe that this form holds in general.

We will now describe how to determine the various $\psi_{p,l}$ terms. This is simpler than it may look, as we do not need to consider the action of the whole $\mathcal{V}$ on $\psi_{p,l}$, but only the action of $\mathcal{V}_l$. Given in fact the structure of (\ref{eq::formSolutionAlgorithm}) this super-operator can act non-trivially only on $\psi_{p,l}$ and its action is particularly simple, as all the terms appearing in $\psi_{p,l}$ have $i_l=j_l$. This means that, in order to deduce the form of $\psi_{p,l}$, we only need to consider those terms in $\beta_p^{(k,k-l)}$ such that $i_l\neq j_l$. 

For the sake of clarity we will describe the method for a specific example. We will hence consider the case for $\theta=\frac{\pi}{12}$ and $\phi=0$, where no resonance occurs. In general, the analysis for any given resonance point (where total domain wall angle is conserved), will depend upon the particular combinations of domain walls that happen to have the same energy. The following analysis needs therefore to be changed accordingly on a case by case basis and works only when we are not at a resonant point. Our purpose is to present a general approach to constructing solutions for $\psi_{p,l}$. There is no real difference in considering solutions for $\psi_{p,l}$ for different $l$s. 

Consider therefore, as an example, the case of $\psi_p^{(5)}$, and suppose that we already know the solution for $\psi_{p,5}$. As we described we can find $\beta_p^{(5,1)}$ from $\beta_p^{(5)}$ and  $\psi_{p,4}$ will be given as
\begin{equation}
    \psi_{p,4}=\sum_{i,j}(\psi_4)_{i_1,j_2,j_3,i_4}^{i_1,i_2,i_3,i_4}|i_1,i_2,i_3,i_4\rangle|i_1,j_2,j_3,i_4\rangle \quad E^{i_1,j_2,j_3,i_4}_{i_1,j_2,j_3,i_4}=0
\end{equation}
and 
\begin{equation}
    \beta_p^{(4,1)}=\sum_{j, i}(\beta_1)_{i_1,j_2,j_3,j_4,i_5}^{i_1,i_2,i_3,i_4,i_5}|i_1,j_2,j_3,j_4,i_5\rangle|i_1,i_2,i_3,i_4,i_5\rangle \quad E^{i_1,i_2,i_3,i_4,i_5}_{i_1,j_2,j_3,j_4,i_5}=0
\end{equation}
As we said we can reconstruct $\psi_{p,4}$ from $\beta_p^{(4,1)}$ by considering the action of $\mathcal{V}_4$ only. This means that we need to look at the elements of $\beta_p^{(4,1)}$ such that $i_4\neq j_4$. We have to distinguish two subcases with respect of the values taken by $i_3$ and $j_3$.

\subsection{Case $i_3\neq j_3$}

To explain the strategy in this situation we consider a specific example. Consider therefore the terms $$(\beta_1)^{0,0,2,0,1}_{0,1,0,1,1}\quad (\beta_1)^{0,0,2,0,2}_{0,1,0,2,2}\quad (\beta_1)^{0,0,2,1,2}_{0,1,0,2,2}\quad (\beta_1)^{0,0,2,2,0}_{0,1,0,0,0}\ ,$$ that is, vectors that agree up to the third indices. Using Mathematica it can be checked that
\begin{equation*}
    \begin{split}
        &\beta^{0,0,2,0,1}_{0,1,0,1,1}=\beta^{0,0,2,0,1}_{0,1,0,2,2}=-\left(\frac{62+71i-(5+60 i) \sqrt{3}}{16 \sqrt{2+\sqrt{3}}}\right)\\
        &\beta^{0,0,2,1,2}_{0,1,0,0,2}=\beta^{0,0,2,2,0}_{0,1,0,0,0}=\left(\frac{62+71i-(5+60 i) \sqrt{3}}{16 \sqrt{2+\sqrt{3}}}\right)\ .
    \end{split}
\end{equation*}
Note that all the coefficients are the same up to a sign. This is in agreement with the fact that they stem from the action of $\mathcal{V}_4$ on a single operator. In order to find what it is, it suffices to consider the operators that we obtain when we make the fourth indices agree by changing only one of them and discarding the fifth indices. For example for the first vector, $|0,0,2,0,1\rangle|0,1,0,1,1\rangle$, we would get the two terms
\begin{equation}
    |0,0,2,1\rangle|0,1,0,1\rangle\quad |0,0,2,0\rangle|0,1,0,0\rangle
\end{equation}
It can be checked that only $|0,0,2,0\rangle|0,1,0,0\rangle$ belongs to $Null(\mathcal{H}_0)$ and therefore this is the only vector that can belong to $\psi_{p,4}$. With some effort it can be shown that when $i_3\neq j_3$ and we are not at a resonance there is always one and only one vector that belongs to $Null(\mathcal{H}_0)$ for any given combination of indexes $i_1,$ $j_1$, etc...~that are not involved\footnote{Note that there exist operators such that by making the $4^{\mathrm{th}}$ indices agree and discarding the last ones, none of the resulting operators belong to $Null(\mathcal{H}_0)$. Consider for example $|0,2,2,2,0\rangle|0,1,1,1,0\rangle$. It can be easily checked then both $|0,2,2,1\rangle|0,1,1,1\rangle$ and $|0,2,2,2\rangle|0,1,1,2\rangle$ are not part of $Null(\mathcal{H}_0)$. 
\\This means that $(\beta_1)^{0,2,2,2,0}_{0,1,1,1,0}=0$, otherwise it would not be possible to find a solution. This is part of the peculiar cancellations that take place when solving these equations.}. Therefore all these vectors can be obtained from the action of $\mathcal{P}_0\mathcal{V}_4$ on
\begin{equation}
    -\left(\frac{62+71i-(5+60 i) \sqrt{3}}{16 \sqrt{2+\sqrt{3}}}\right)|0,0,2,0\rangle|0,1,0,0\rangle\otimes I_1\ .
\end{equation}
In other words $(\psi_4)^{0,0,2,0}_{0,1,0,0}=-\left(\frac{62+71i-(5+60 i) \sqrt{3}}{16 \sqrt{2+\sqrt{3}}}\right)$. Note that all this is possible because the coefficients for $\beta$ follow the symmetries that we have highlighted above. This procedure can then be repeated for all the terms in $\beta^{(4,1)}$ such that $i_4\neq j_4$ and $i_3\neq j_3$. We can now consider the case of $i_3=j_3$.

\subsection{Case $i_3=j_3$}

Also in this case, to understand the point, it is better to consider a specific case. Take therefore the terms
\begin{equation*}
\begin{split}
&(\beta_1)^{0,0,2,2,0}_{0,2,2,0,0}\quad  (\beta_1)^{0,0,2,1,1}_{0,2,2,2,1}\quad(\beta_1)^{0,0,2,0,2}_{0,2,2,1,2}\\ &(\beta_1)^{0,0,2,1,2}_{0,2,2,0,2}\quad (\beta_1)^{0,0,2,0,0}_{0,2,2,2,0}\quad (\beta_1)^{0,0,2,2,1}_{0,2,2,1,1}\ .
\end{split}    
\end{equation*}
It can be checked with Mathematica that
\begin{equation*}
    \begin{split}
        (\beta_1)^{0,0,2,2,1}_{0,2,2,1,1}=-(\beta_1)^{0,0,2,1,1}_{0,2,2,2,1}\quad(\beta_1)^{0,0,2,0,0}_{0,2,2,2,0}=-(\beta_1)^{0,0,2,2,0}_{0,2,2,0,0}\quad(\beta_1)^{0,0,2,1,2}_{0,2,2,0,2}=-(\beta_1)^{0,0,2,0,1}_{0,2,2,1,0}
    \end{split}
\end{equation*}
and
\begin{equation}
    \begin{split}
        &(\beta_1)^{0,0,2,2,0}_{0,2,2,0,0}=-\frac{-276 + 117i + 217\sqrt{3}}{
  8\sqrt{2 + \sqrt{3}} (-37 - 14 i + (20 + 3 i) \sqrt{3}))} + 
 \frac{1}{52}\sqrt{1980 - 2337i}\\
 &(\beta_1)^{0,0,2,1,1}_{0,2,2,2,1}=- \frac{1}{52}\sqrt{1980 - 2337i}\\
 &(\beta_1)^{0,0,2,0,2}_{0,2,2,1,2}=\frac{-276 + 117i + 217\sqrt{3}}{
  8\sqrt{2 + \sqrt{3}} (-37 - 14 i + (20 + 3 i) \sqrt{3}))}
    \end{split}
\end{equation}
Note that 
\begin{equation}
\label{eq::conditionAlgorithm1}
    (\beta_1)^{0,0,2,2,0}_{0,2,2,0,0}+(\beta_1)^{0,0,2,1,1}_{0,2,2,2,1}+(\beta_1)^{0,0,2,0,2}_{0,2,2,1,2}=0\ .
\end{equation}
Again we need to consider the action of $\mathcal{P}_0\mathcal{V}_4$ on vectors of length 4. That is
\begin{equation*}
    \small
    \begin{split}
   (\psi_4)^{0,0,2,0}_{0,2,2,0}|0,0,2,0\rangle|0,2,2,0\rangle+(\psi_4)^{0,0,2,1}_{0,2,2,1}|0,0,2,1\rangle|0,2,2,1\rangle+(\psi_4)^{0,0,2,2}_{0,2,2,2}|0,0,2,2\rangle|0,2,2,2\rangle\ .
   \end{split}
\end{equation*}
Acting with $\mathcal{P}_0\mathcal{V}_4$ on this operator (tensored with $I_1$) and equating it with the terms in $\beta_p^{(4,1)}$ that we are interested in, yields the system of equations
\begin{equation}
    \begin{cases}
    &(\psi_4)^{0,0,2,2}_{0,2,2,2}-(\psi_4)^{0,0,2,0}_{0,2,2,0}=(\beta_1)^{0,0,2,2,0}_{0,2,2,0,0}\\
    &(\psi_4)^{0,0,2,1}_{0,2,2,1}-(\psi_4)^{0,0,2,2}_{0,2,2,2}=(\beta_1)^{0,0,2,1,1}_{0,2,2,2,1}\\
    &(\psi_4)^{0,0,2,0}_{0,2,2,0}-(\psi_4)^{0,0,2,1}_{0,2,2,1}=(\beta_1)^{0,0,2,0,2}_{0,2,2,1,2}\ .
    \end{cases}
\end{equation}
Since condition (\ref{eq::conditionAlgorithm1}) holds, the system admits solutions and we can set 
\begin{equation}
    \begin{split}
        &(\psi_4)^{0,0,2,0}_{0,2,2,0}=\frac{\beta^{0,0,2,0,2}_{0,2,2,1,2}-\beta^{0,0,2,2,0}_{0,2,2,0,0}}{3}+\lambda\\
        &(\psi_4)^{0,0,2,1}_{0,2,2,1}=\frac{\beta^{0,0,2,1,1}_{0,2,2,2,1}-\beta^{0,0,2,0,2}_{0,2,2,1,2}}{3}+\lambda\\
        &(\psi_4)^{0,0,2,2}_{0,2,2,2}=\frac{\beta^{0,0,2,2,0}_{0,2,2,0,0}-\beta^{0,0,2,1,1}_{0,2,2,2,1}}{3}+\lambda
    \end{split}
\end{equation}
with some arbitrary constant $\lambda$ that we can take to be $0$. The value that we chose for this constant is not important as we know from our discussion in Section \ref{sec::RestrictionOnTheSolution} that a basis for the solution of
\begin{equation}
    \mathcal{P}_0\mathcal{V}(\psi_p^{(5)}\otimes I_1)=0
\end{equation}
is given by $\sigma_1$. This means that at the end of this process, after we determined all the $\psi_{p,l}$, the various arbitrary constants that we fix during the solution will reduce to some multiple of $\sigma_1$ (always assuming that the algorithm will produce a legitimate solution of the problem)\footnote{As we saw in Section \ref{sec::Normalization}, we can exploit the addition of linear combinations of $\sigma_1$ to fix the normalization of the zero mode, so at this level it does not really matter which constant multiplies $\sigma_1$.}. 

We can repeat this procedure for all terms of $\beta_p^{(4,1)}$ such that $i_3=j_3$ and then continue to the next step where we consider $\beta_p^{(4,2)}$ as described. The process can now be repeated in the same way again by considering terms with $i_3\neq j_3$, then considering the subcases $i_2\neq j_2$, $i_2=j_2$ and so on until we find all the terms in $\psi_p^{(k)}$.

This concludes our explanation on how to find solutions for $\psi_p$. Although the details given here only apply to the case of the chiral Potts model, the procedure ultimately reduces to finding the action of the pseudo-inverse of $\mathcal{P}_0\mathcal{V}_l\mathcal{P}_0$ for the various $l\leq k$. It can therefore be generalized to consider the other models that we mentioned in Section \ref{sec::GeneralFormOfTheSolution} as well.

\section{Conclusions}
\label{sec::Conclusions}

We have analysed the construction of strong zero modes in $\mathbb{Z}_N$ parafermionic chain models. Although we applied our method specifically to the $\mathbb{Z}_3$ case the techniques that we have shown here can be generalised to all prime $N$ and to many other spin chain models. 

We investigated in particular the connection of the existence of parafermionic zero modes with the conservation of the total domain wall angle, and the iterative constuction of zero mode operators, generalizing results from \cite{Fendley2012,XYZ,Kemp_2017}. In addition we showed that the existence of zero modes is connected to a perturbation theory problem at the level of super-operators. More precisely, this is a constrained perturbative problem in the degenerate null space of the commutator with the free Hamiltonian.

We have shown directly that, at resonance points where the total domain wall angle is not conserved, local zero modes cannot be constructed and we have demonstrated that the conservation of total domain wall angle amounts to a condition regarding the locality of the terms appearing in the iterative expansion. Ultimately we find that this last property is what allows the construction of zero modes in this and similar models.

We also addressed the problem of normalization of the zero mode in the thermodynamic limit and we have shown that properties of the zero mode at $f=0$ (e.g. symmetries) generalize to $f\neq 0$ and this allows us to show how a normalized zero mode can be constructed at all values of $L$.
However the problem of convergence of the expansion is still largely unanswered (this is connected to the problems of prethermalization and long coherence time for spin edges \cite{Kemp_2017,Else2017,Vasiloiu2019}).

We have provided a general ansatz for the shape taken by the solution of the perturbative expansion. These results have been checked using numerical symbolic calculations up to $5^{\mathrm{th}}$ order, even though we where able to find a solution up $6^{\mathrm{th}}$ order. However we were not able to check if this last solution is consistent with the provided ansatz, although there are strong suggestions that it does. We find that this framework agrees with other known iterative constructions of strong zero modes \cite{XYZ,Kemp_2017}. This raises questions about the reasons that allow such methods and the accompanying remarkable cancellations to work. Further developments in this direction are needed in order to fully understand this.

\section{Acknowledgments}
The authors would like to thank Stephen Nulty, Kevin Kavanagh and Aaron Conlon for useful discussions and comments.  D.P. and J.K.S. acknowledge financial support from Science Foundation Ireland through Principal Investigator Awards 12/IA/1697 and 16/IA/4524. G.K. acknowledges support from Science Foundation Ireland through Career Development Award 15/CDA/3240.

\bibliographystyle{unsrt}
\bibliography{Bibliography}

\appendix

\section{Proof of Property 2 and 3}
\label{appendix::ProofProperty24}
In this appendix we will prove the Properties given in Section \ref{sec::RestrictionOnTheSolution}. We will first consider Property 2
\newline
\\\textbf{Property 2.} \textit{Consider the equation
\begin{equation}
\label{eq::AppendixNecessaryCondition}
    \mathcal{P}_0\mathcal{V}\alpha_p\otimes I_1=\beta_p\otimes I_1
\end{equation}
with $\alpha_p,\beta_p\in Null(\mathcal{H}_0)$ given as
\begin{equation}
\label{eq::AppendixFormAlphap}
    \alpha_p=\sum\alpha^{\mathbf{i}_{t'-1},i_{t'}}_{\mathbf{j}_{t-1},i_{t'}}|\mathbf{i}_{t'-1},i_{t'}\rangle|\mathbf{j}_{t'-1},i_{t'}\rangle
\end{equation}
and
\begin{equation*}
\label{eq::AppendixFormBetap}
    \beta_p=\sum\beta^{\mathbf{i}_{t-1},i_t}_{\mathbf{j}_{t-1},i_t}|\mathbf{i}_{t-1},i_t\rangle|\mathbf{j}_{t-1},i_t\rangle\ .
\end{equation*}
Assuming such equation to hold then, necessarily, $t'=t$ and
\begin{equation}
    \alpha_p=\sum\alpha^{\mathbf{i}_{t-2},i_{t-1}}_{\mathbf{j}_{t-2},i_{t-1}}|\mathbf{i}_{t-2},i_{t-1}\rangle|\mathbf{j}_{t-2},i_{t-1}\rangle\otimes I_1\ .
\end{equation}}
\begin{proof}
To prove this Property consider the action of $\mathcal{V}$ on $\alpha_p$. We have
\begin{equation}
    \mathcal{V}\alpha_p=\sum_{k=1}^{t'-1}\mathcal{V}_k\alpha_p+\mathcal{V}_{t'}\alpha_p\ .
\end{equation}
Since the last two left and right indices in (\ref{eq::AppendixFormAlphap}) are the same for all the terms in the sum, we can easily write down the action of $\mathcal{V}_{t'}$ on $\alpha_p$. That is
\begin{equation}
\label{eq::Valphap}
\begin{split}
    \mathcal{V}_{t'}\alpha_p&=\sum\left(\alpha^{\mathbf{i}_{t'-1},i_{t'}+1}_{\mathbf{j}_{t'-1},i_{t'}+1}-\alpha^{\mathbf{i}_{t'-1},i_{t'}}_{\mathbf{j}_{t'-1},i_{t'}}\right)e^{-i\phi}|\mathbf{i}_{t'-1},i_{t'}+1\rangle|\mathbf{j}_{t'-1},i_{t'}\rangle\\
    &\ \   +\sum\left(\alpha^{\mathbf{i}_{t'-1},i_{t'}-1}_{\mathbf{j}_{t'-1},i_{t'}-1}-\alpha^{\mathbf{i}_{t'-1},i_{t'}}_{\mathbf{j}_{t'-1},i_{t'}}\right)e^{i\phi}|\mathbf{i}_{t'-1},i_{t'}-1\rangle|\mathbf{j}_{t'-1},i_{t'}\rangle
    \ .
\end{split}
\end{equation}
In order to consider solutions of (\ref{eq::AppendixNecessaryCondition}) we now need to project down these terms in $Null(\mathcal{H}_0)$. As we saw, since $\mathcal{H}_0$ can act on two sites per time, we need to consider the addition of an identity operator at the end of the chain. This means that we need to consider operators of the type
\begin{equation}
\label{eq::necessaryConditionSolution}
    |\mathbf{i}_{t'-1},i_{t'}\pm1,i_{t'+1}\rangle|\mathbf{j}_{t'-1},i_{t'},i_{t'+1}\rangle\ .
\end{equation}
Projecting down into $Null(\mathcal{H}_0)$ consists on finding values $i_{t'+1}\in\mathbb{Z}_3$ such that
\begin{equation}
  E^{\mathbf{i}_{t'-1},i_{t'}\pm1,i_{t'+1}}_{\mathbf{j}_{t'-1},\ \ i_{t'}\ ,i_{t'+1}}=0\ .
\end{equation}
To simplify the problem we can write $E^{\mathbf{i}_{t'-1},i_{t'}\pm1,i_{t'+1}}_{\mathbf{j}_{t'-1},\ \ i_{t'}\ ,i_{t'+1}}$ as
\begin{equation*}
  E^{\mathbf{i}_{t'-1},i_{t'}\pm1,i_{t'+1}}_{\mathbf{j}_{t'-1},\ \ i_{t'}\ ,i_{t'+1}}=E^{\mathbf{i}_{t'-1},i_{t'}}_{\mathbf{j}_{t'-1},i_{t'}}+E^{i_{t'-1},i_{t'}\pm1,i_{t'+1}}_{i_{t'-1},\ \ i_{t'}\ ,i_{t'+1}}\ ,
\end{equation*}
where we have gathered all the final domain walls in the second term. 
Since by hypothesis \mbox{$E^{\mathbf{i}_{t'-1},i_{t'}}_{\mathbf{j}_{t'-1},i_{t'}}=0$}, we are left with
\begin{equation*}
  E^{\mathbf{i}_{t'-1},i_{t'}\pm1,i_{t'+1}}_{\mathbf{j}_{t'-1},\ \ i_{t'}\ ,i_{t'+1}}=E^{i_{t'-1},i_{t'}\pm1,i_{t'+1}}_{i_{t'-1},\ \ i_{t'}\ ,i_{t'+1}}\ .
\end{equation*}
Consider now $E^{i_{t'-1},i_{t'}\pm1,i_{t'+1}}_{i_{t'-1},\ \ i_{t'}\ ,i_{t'+1}}$. In terms of domain wall energies we have
\begin{equation*}
    E^{\mathbf{i}_{t'-1},i_{t'}\pm1,i_{t'+1}}_{\mathbf{j}_{t'-1},\ \ i_{t'}\ ,i_{t'+1}}=\epsilon_{i_{t'+1}-i_{t'}}+\epsilon_{i_{t'}-i_{t'-1}}-\epsilon_{i_{t'+1}-i_{t'}\mp1}-\epsilon_{i_{t'}\pm1-i_{t'-1}}\ ,
\end{equation*}
with $\epsilon_m$ as in (\ref{eq::domainWallEnergies}). If we now set
\begin{equation}
    i_{t'+1}=2i_{t'}\pm1-i_{t'-1} \mod 3
\end{equation}
we get
\begin{equation}
    E^{\mathbf{i}_{t'-1},i_{t'}\pm1,i_{t'+1}}_{\mathbf{j}_{t'-1},\ \ i_{t'}\ ,i_{t'+1}}=0
\end{equation}
This means that all the terms in (\ref{eq::Valphap}) will always produce operators that belong to the $Null(\mathcal{H}_0)$ when we tensor them out with $I_1$. Therefore the operator $\mathcal{P}_0\mathcal{V}_t\alpha_p\otimes I_1$ will always invariably contain terms of the form
\begin{equation}
\label{eq::termsAlphap}
     |\mathbf{i}_{t'-1},i_{t'},i_{t'+1}\rangle|\mathbf{j}_{t'-1},j_{t'},i_{t'+1}\rangle\qquad i_{t'}\neq j_{t'}\ ,
\end{equation}
that cannot be removed from the rest of the action of $\mathcal{V}$, since they leave the last left-right index the same.
Imposing that $\alpha_p$ constitutes a solution of (\ref{eq::necessaryConditions}), with $\beta_p$ as in (\ref{eq::necessaryConditionBeta}) means that $t=t'$ and that
\begin{equation}
\label{eq::Vtcondition}
\mathcal{V}_t\alpha_p=0\ ,
\end{equation}
because in $\beta_p$ all the terms have the same left-right indices at site t, while we saw that if (\ref{eq::Vtcondition}) is not satisfied then $\mathcal{V}_t$ will produces terms of the form (\ref{eq::termsAlphap}) when acts on $\alpha_p$. Condition (\ref{eq::Vtcondition}), in turn means that
\begin{equation}
    \alpha_p=\sum\alpha^{\mathbf{i}_{t-2},i_{t-1}}_{\mathbf{j}_{t-2},i_{t-1}}|\mathbf{i}_{t-2},i_{t-1}\rangle|\mathbf{j}_{t-2},i_{t-1}\rangle\otimes I_1\ ,
\end{equation}
which proves the statement.
\end{proof}

We can now consider the other Property given in Section \ref{sec::RestrictionOnTheSolution}
\newline
\\\textbf{Property 3.} \textit{
When $\theta\neq0,\frac{\pi}{3},\frac{2\pi}{3}$, the unique possible local starting point of the zero mode expansion is given by
\begin{equation}
    \psi^{(0)}_p=\sigma_1\ .
\end{equation}
When $\theta=0,\frac{\pi}{3},\frac{2\pi}{3}$ a local zero mode cannot exists.}
\begin{proof}
Suppose by contradiction that 
\begin{equation}
\psi^{(0)}_p=\sum\psi^{\mathbf{i}_{t-1},i_{t}}_{\mathbf{j}_{t-1},i_{t}}|\mathbf{i}_{t-1},i_{t}\rangle|\mathbf{j}_{t-1},i_{t}\rangle
\end{equation}
for some $t\in\mathbb{N}^+$. In order to be a good starting point of our zero mode expansion $\psi^{(0)}_p$ has to satisfy
\begin{equation}
    \mathcal{P}_0\mathcal{V}(\psi^{(0)}\otimes I_1)=0\ .
\end{equation}
As we saw in the proof of Property \ref{property::GeneralConditionAlphaBeta}, the action of $\mathcal{P}_0\mathcal{V}$ on $\psi_p^{(0)}\otimes I_1$ will always be different from zero when it acts on the firs index, unless
\begin{equation}
    \mathcal{V}_1\psi_p^{(0)}=0\ .
\end{equation}
By induction this means that 
\begin{equation}
    \psi_p^{(0)}=\sum_{i}\psi^i_i|i\rangle|i\rangle\ .
\end{equation}
As we saw in the main text, if $\theta=0,\frac{\pi}{3},\frac{2\pi}{3}$, then the only possible solution is given by $\psi_p^{(0)}\propto I_1$. This solution is inconsistent with the condition 
\begin{equation}
    \mathcal{Q}\psi_p^{(0)}=\omega\psi_p^{(0)}\ ,
\end{equation}
since
\begin{equation}
     \mathcal{Q}I_1= I_1\ .
\end{equation}
Thus, if $\theta=0,\frac{\pi}{3},\frac{2\pi}{3}$, there is no local solution to the zero mode expansion problem. If $\theta$ is not one of this resonant points than the solution, up to a multiplicative constant, is given by
\begin{equation}
   \psi_p^{(0)}= \sigma_1\ .
\end{equation}
To directly check all this and see how the total domain wall angle conservation enters into the problem, consider 
\begin{equation*}
    \mathcal{V}\psi_p^{(0)}\otimes I_1=(-e^{i\phi}\tau_1^L+e^{-i\phi}\tau_1^R+h.c)\psi_p^{(0)}\otimes I_1\ .
\end{equation*}
To compute this expression consider
\begin{equation*}
    (-e^{i\phi}\tau_1^L+e^{-i\phi}\tau_1^R+h.c)|0\rangle|0\rangle=-e^{i\phi}|2\rangle|0\rangle+e^{-i\phi}|0\rangle|2\rangle-e^{-i\phi}|1\rangle|0\rangle+e^{i\phi}|0\rangle|1\rangle\ .
\end{equation*}
As we saw in Section \ref{sec::Symmetry} we can use the symmetry under $\mathcal{Q}$ and $\mathcal{T}\mathcal{K}$ to build the complete action of $\mathcal{V}$ on $\psi_p^{(0)}$ and this yelds
\begin{equation}
\label{eq::Appendixtemp1}
    \mathcal{V}\psi_p^{(0)}\otimes I_1=-\sum_{i_1\neq j_1,i_2} (\omega^{j_1}-\omega^{i_1})e^{i(j_1-i_1)\phi}|i_1,i_2\rangle|j_1,i_2\rangle\ .
\end{equation}
Note that the difference between the two indexes $j_1-i_1$ has been introduced for notation convenience and has to be intended as taking values in $\{-1,1\}$. This means that in the event it results equal to $2$ or $-2$, it has to be substituted with $-1$ or $1$ respectively. We will keep using the same convention throughout the rest of the appendixes.
Since we are supposing that total domain wall angle is conserved then it follows that
\begin{equation}
\label{eq::Appendixtemp0}
    \mathcal{P}_0\mathcal{V}\psi_p^{(0)}\otimes I_1=0\ . 
\end{equation}
because $i_1\neq j_1$ in (\ref{eq::Appendixtemp1}), while operators in $Null(\mathcal{H}_0)$ have $i_1=j_1$ as prescribed by Property \ref{property::Null}.
\end{proof}

\section{Proof of Property 4 and Property 5}
\label{appendix:ProofProperty5}

In this appendix we will prove Property \ref{property::conditionExistence}, that we recall here for convenience
\newline
\\\textbf{Property 4.} \textit{
Suppose that a local solution for the zero mode expansions exists up to some order $k$. If the total domain wall angle is conserved then
\begin{equation}
\label{eq::AppendixconditionExistence}
     \mathcal{P}_0(\mathcal{V}\psi_q^{(k)}\otimes I_1)=\beta_p^{(k)}\otimes I_1\ .
\end{equation}
with $\beta_p^{(k)}\in Null(\mathcal{})_0$
}

This property haHs to do with the general structure of the solution, in the following we will prove a generalization of it
\newline
\\\textbf{Property 4'.} \textit{
Suppose that a solution for the zero mode exists up to some order $k\geq 1$. If the total domain wall angle is conserved then
\begin{equation*}
     \mathcal{P}_0(\mathcal{V}\psi_q^{(k)}\otimes I_1)=\beta_p^{(k)}\otimes I_1\ .
\end{equation*}
with $\beta_p^{(k)}\in Null(\mathcal{H}_0)$ and the solution $\psi^{(k)}$ is constituted by operators that act non trivially on chain of length $k+1$. In particular
\begin{itemize}
    \item The operators $\psi_q^{(k)}$ can be written as 
    \begin{equation*}
        \psi_q^{(k)}=\eta_q^{(k)}+\rho_q^{(k)}
    \end{equation*}
    where $\eta_q^{(k)}$ is a linear combination of operators of the form
    \begin{equation}
\label{eq::basisChiTilde}
    |i_1,i_2,\ldots,i_k,i_{k+1}\rangle|j_1,j_2,\ldots,j_k,i_{k+1}\rangle\qquad i_l\neq j_l \qquad \forall\ l=1,2\ldots,k
\end{equation}
   while $\rho_q^{(k)}$ acts trivially on site $k+1$ and is a linear combination of operators of the form
    \begin{equation*}
         |\mathbf{i}_{k-1},i_k\rangle|\mathbf{j}_{k-1},i_k\rangle\otimes I_1\qquad E^{\mathbf{i}_{k-1},i_k}_{\mathbf{j}_{k-1},i_k}\neq0\ .
    \end{equation*}
     \item $\psi_p^{(k)}$ is made of operators that act non-trivially on chains that are not longer than $k$. This means that $\psi_p^{(k)}$ is a linear combinations of operators of the form
    \begin{equation*}
        |i_1,\mathbf{i}_{k-2},i_k\rangle|i_1,\mathbf{j}_{k-2},i_k\rangle \otimes I_1\qquad E^{i_1,\mathbf{i}_{k-2},i_k}_{i_1,\mathbf{j}_{k-2},i_k}=0\ .
    \end{equation*}
\end{itemize}}
\begin{proof}
We will first prove that 
\begin{equation}
        \psi_q^{(k)}=\eta_q^{(k)}+\rho_q^{(k)}
\end{equation}
by induction on $j\leq k$. 
\newline
\\\textbf{Case j=1}
\newline
\\We start by considering $\psi^{(0)}$. We have 
\begin{equation}
    \psi^{(0)}=\psi_p^{(0)}=\sum_{i}\omega^i|i\rangle|i\rangle\ .
\end{equation}
As we saw in Appendix \ref{appendix::ProofProperty24}, if the total domain wall angle is conserved, then 
\begin{equation}
    \mathcal{P}_0\mathcal{V}\psi_p^{(0)}\otimes I_1=0\ . 
\end{equation}
and 
\begin{equation}
    \mathcal{V}\psi_p^{(0)}\otimes I_1=-\sum_{i_1\neq j_1,i_2} (\omega^{j_1}-\omega^{i_1})e^{i(j_1-i_1)\phi}|i_1,i_2\rangle|j_1,i_2\rangle\ .
\end{equation}
In order to find $\psi^{(1)}$ we now need to use equations (\ref{eq::BrillouinWigner1Identity}), that in the present case are given by
\begin{equation}
\label{eq::AppendixDefiningEquationPsi1}
    \begin{split}
        &\mathcal{Q}_0\mathcal{H}_0\psi_q^{(1)}=-\mathcal{Q}_0\mathcal{V}(\psi_q^{(1)}\otimes I_1)\\
        &\mathcal{P}_0\mathcal{V}(\psi_p^{(1)}\otimes I_1)=-\mathcal{P}_0\mathcal{V}(\psi_q^{(1)}\otimes I_1)\ .
    \end{split}
\end{equation}
As we saw the first of these equations is easily solved:
\begin{equation*}
    \psi^{(1)}_q=-\frac{\mathcal{Q}_0}{\mathcal{H}_0}\mathcal{V}(\psi_p^{(0)}\otimes I_1)=\sum_{i_1\neq j_1,i_2} (\omega^{j_1}-\omega^{i_1})\frac{e^{i(j_1-i_1)\phi}}{E^{i_1,i_2}_{j_1,i_2}}|i_1,i_2\rangle|j_1,i_2\rangle\ .
\end{equation*}

In order to find a solution for $\psi_p^{(1)}$ we need to repeat what we did for $\psi^{(0)}$:
\begin{equation*}
    \begin{split}
        \mathcal{V}\psi_q^{(1)}\otimes I_1&=-\sum_{i_1\neq j_1,i_2\neq j_2,i_3}(\omega^{j_1}-\omega^{i_1})\left(\frac{1}{E^{i_1,j_2}_{j_1,j_2}}-\frac{1}{E^{i_1,i_2}_{j_1,i_2}}\right)e^{i(j_1-i_1)\phi}e^{i(j_2-i_2)\phi}|i_1,i_2,i_3\rangle|j_1,j_2,i_3\rangle\\
        &-\sum_{i_1\neq j_1,i_2,i_3}(\omega^{j_1}-\omega^{i_1})\left(\frac{e^{2i(j_1-i_1)\phi}}{E^{2i_1-j_1,i_2}_{\phantom{2i_1}j_1\phantom{-},i_2}}-\frac{e^{-2i(j_1-i_1)\phi}}{E^{\phantom{2j_1}i_1\phantom{-},i_2}_{2j_1-i_1,i_2}}\right)e^{i(i_1-j_1)\phi}|i_1,i_2,i_3\rangle|j_1,i_2,i_3\rangle\ .
    \end{split}
\end{equation*}
Where the terms with $i_1=j_1$ cancel identically and are therefore excluded from the sum. Since all terms in the sum have $i_1\neq j_1$, if the total domain wall angle is conserved\footnote{For chains of length $L=3$ the resonance points are given by $\theta=0,\ \frac{\pi}{3},\ \frac{2\pi}{3}$ and $\theta=\frac{\pi}{6}$. Since the resonance at $\theta=\frac{\pi}{6}$ conserves total domain wall angle the only problematic phases are $\theta=0,\ \frac{\pi}{3},\ \frac{2\pi}{3}$, as chains for length $L=2$.}, we have
\begin{equation}
\label{eq::Appendixtemp2}
    \mathcal{P}_0\mathcal{V}(\psi^{(1)}_q\otimes I_1)=0\ .
\end{equation}
Therefore in the case j=1
\begin{equation}
    \mathcal{P}_0\mathcal{V}(\psi^{(1)}_q\otimes I_1)= \beta_p^{(1)}\otimes I_1\ ,
\end{equation}
with $\beta_p^{(1)}=0\in Null(\mathcal{H}_0)$. We therefore conclude that $\psi_p^{(1)}=0$ constitute a solution of (\ref{eq::AppendixDefiningEquationPsi1})\footnote{It can be proved that this solution also satisfy the requirement on the normalization that we discussed in Section \ref{sec::Normalization}.}.
\newline
\\\textbf{Case j=l-1}
\newline
\\Suppose now that Property \ref{property::conditionExistence} holds for $j=l-1$. We have
\begin{equation*}
    \psi_q^{(l)}=-\frac{\mathcal{Q}_0}{\mathcal{H}_0}\mathcal{V}\psi^{(l-1)}\otimes I_1=-\frac{1}{\mathcal{H}_0}\mathcal{V}\left(\eta_q^{(l-1)}+\rho_q^{(l-1)}+\psi_p^{(l-1)}\right)\otimes I_1
\end{equation*}
Consider now $\mathcal{V}$ as a sum of its local terms. Since $\psi^{(l-1)}$ lives on a chain of length $l$ we can consider $\mathcal{V}$ only up to terms that act on site $l$ on the chain
\begin{equation}
    \mathcal{V}=\sum_{t=1}^{l-1}\mathcal{V}_t+\mathcal{V}_l=\widetilde{\mathcal{V}}_{l-1}+\mathcal{V}_l\ ,
\end{equation}
where we defined 
\begin{equation}
    \widetilde{\mathcal{V}}_{l-1}=\sum_{t=1}^{l-1}\mathcal{V}_t\ .
\end{equation}
Note that by hypothesis $\rho_q^{(l-1)}$, $\psi_p^{(l-1)}$ act non trivially on a chain only up to sites $l-1$, which means that the action of $\mathcal{V}_l$ is 0 on them. Hence we have
\begin{equation}
    \psi_q^{(l)}=-\frac{\mathcal{Q}_0}{\mathcal{H}_0}\left(\mathcal{V}_l\eta_q^{(l-1)}+\widetilde{\mathcal{V}}_{l-1}\left(\eta_q^{(l-1)}+\rho_q^{(l-1)}+\psi_p^{(l-1)}\right)\right)\otimes I_1\ .
\end{equation}
Suppose therefore that 
\begin{equation*}
   \mathcal{Q}_0\widetilde{\mathcal{V}}_{l-1}\left(\eta_q^{(l-1)}+\rho_q^{(l-1)}+\psi_p^{(l-1)}\right)\otimes I_1=\sum\beta^{\mathbf{i}_{l-1},i_l}_{\mathbf{j}_{l-1},i_l}|\mathbf{i}_{l-1},i_l\rangle|\mathbf{j}_{l-1},i_l\rangle\otimes I_1\ ,
\end{equation*}
where the operators in the sum have the same indices at site $l$ because $\widetilde{\mathcal{V}}_{l-1}$ acts trivially at site $l$.
Therefore we have
\begin{equation*}
    \begin{split}
        &\frac{\mathcal{Q}_0}{\mathcal{H}_0}\widetilde{\mathcal{V}}_{l-1}\left(\eta_q^{(l-1)}+\rho_q^{(l-1)}+\psi_p^{(l-1)}\right)\otimes I_1=\sum\frac{\beta^{\mathbf{i}_{l-1},i_l}_{\mathbf{j}_{l-1},i_l}}{E^{\mathbf{i}_{l-1},i_l,i_{l+1}}_{\mathbf{j}_{l-1},i_l,i_{l+1}}}|\mathbf{i}_{l-1},i_l,i_{l+1}\rangle|\mathbf{j}_{l-1},i_l,i_{l+1}\rangle\ ,
    \end{split}
\end{equation*}
where the dependence of the coefficients $\beta$ on $i_l$ comes from $\eta_q^{(l-1)}$. Using that $E^{\mathbf{i}_{l-1},i_l,i_{l+1}}_{\mathbf{j}_{l-1},i_l,i_{l+1}}=E^{\mathbf{i}_{l-1},i_l}_{\mathbf{j}_{l-1},i_l}$ we get
\begin{equation*}
\sum\frac{\beta^{\mathbf{i}_{l-1},i_l}_{\mathbf{j}_{l-1},i_l}}{E^{\mathbf{i}_{l-1},i_l}_{\mathbf{j}_{l-1},i_l}}|\mathbf{i}_{l-1},i_l,i_{l+1}\rangle|\mathbf{j}_{l-1},i_l,i_{l+1}\rangle=\sum\frac{\beta^{\mathbf{i}_{l-1},i_l}_{\mathbf{j}_{l-1},i_l}}{E^{\mathbf{i}_{l-1},i_l}_{\mathbf{j}_{l-1},i_l}}|\mathbf{i}_{l-1},i_l\rangle|\mathbf{j}_{l-1},i_l\rangle\otimes I_1\ ,
\end{equation*}
hence we can set
\begin{equation}
    \rho_q^{(l)}=-\frac{\mathcal{Q}_0}{\mathcal{H}_0}\widetilde{\mathcal{V}}_{l-1}\left(\eta_q^{(l-1)}+\rho_q^{(l-1)}+\psi_p^{(l-1)}\right)\otimes I_1\ .
\end{equation}

Suppose now that 
\begin{equation}
    \eta_q^{(l-1)}=\sum_{i_t\neq j_t}\eta^{i_1,i_2,\ldots,i_{l-1},i_{l}}_{j_1,j_1,\ldots,j_{l-1},i_{l}} |i_1,i_2,\ldots,i_{l-1},i_{l}\rangle|j_1,j_2,\ldots,j_{l-1},i_{l}\rangle\ .
\end{equation}
the action of $\mathcal{V}_l$ is then given by:
\begin{equation*}
    \begin{split}
        \mathcal{V}_l\eta_q^{(l-1)}=\sum_{i_t\neq j_t}\left(\eta^{i_1,i_2,\ldots,i_{l-1},i_{l}}_{j_1,j_1,\ldots,j_{l-1},i_{l}}-\eta^{i_1,i_2,\ldots,i_{l-1},j_{l}}_{j_1,j_1,\ldots,j_{l-1},j_{l}}\right)e^{i(j_l-i_l)\phi} |i_1,i_2,\ldots,i_{l-1},i_{l}\rangle|j_1,j_2,\ldots,j_{l-1},j_{l}\rangle\ .
    \end{split}
\end{equation*}
Since the left and right $l$ indices are different from each other we can write
\begin{equation}
    \eta_q^{(l)}=-\frac{\mathcal{Q}_0}{\mathcal{H}_0}(\mathcal{V}_l\eta_q^{(l-1)}\otimes I_1)\ .
\end{equation}
This proves the first part of the property, we can now prove that
 \begin{equation*}
        |i_1,\mathbf{i}_{k-2},i_k\rangle|i_1,\mathbf{j}_{k-2},i_k\rangle \otimes I_1\qquad E^{i_1,\mathbf{i}_{k-2},i_k}_{i_1,\mathbf{j}_{k-2},i_k}=0
    \end{equation*}
constitute a basis for $\psi_p^{(l)}$. Again we recall Property (\ref{eq::propertyEnergy}), which states that whenever we have the same domain wall at end of the chain, on the left and on the right sector, they cancel out when computing the energy
\begin{equation}
\label{eq::propertyEnergy1}
    E_{\mathbf{i}_{l-1},i_l,i_{l+1}}^{\mathbf{j}_{l-1},i_l,i_{l+1}}=E_{\mathbf{i}_{l-1},i_{l}}^{\mathbf{j}_{l-1},i_{l}}\qquad \forall\ i_l=0,1,2\ .
\end{equation}
This means in particular that
\begin{equation}
    \mathcal{P}_0(\mathcal{V}|\mathbf{i}_{l-1},i_l\rangle|\mathbf{j}_{l-1},i_l\rangle \otimes I_2)=(\mathcal{P}_0\mathcal{V}|\mathbf{i}_{l-1},i_l\rangle|\mathbf{j}_{l-1},i_l\rangle \otimes I_1)\otimes I_1\ ,
\end{equation}
As can be explicitly checked. Therefore, extending by linearity, we have
\begin{equation}
\mathcal{P}_0(\mathcal{V}\psi_q^{(l)}\otimes I_1)=(\mathcal{P}_0\mathcal{V}\rho_q^{(l)})\otimes I_1\ .    
\end{equation}
Thus the only problems when we consider (\ref{eq::conditionExistence}), may arise from $\eta_q^{(l)}$. If the total domain wall angle is conserved, however, this can never happen. In (\ref{eq::basisChiTilde}), in fact, $i_1\neq j_1$ and we know that operators in $Null(\mathcal{H}_0)$ need to have $i_1=j_1$ because of Property \ref{property::Null}. This means that the action of $\mathcal{V}_1$ is the only one that can survive after the projection into $Null(\mathcal{H}_0)$:
\begin{equation*}
    \mathcal{P}_0(\mathcal{V}\eta_q^{(l)}\otimes I_1)=\mathcal{P}_0(\mathcal{V}_1\eta_q^{(l)}\otimes I_1)+\mathcal{P}_0(\sum_{t=2}^{l+1}\mathcal{V}_t\eta_q^{(l)}\otimes I_1)=\mathcal{P}_0(\mathcal{V}_1\eta_q^{(l)}\otimes I_1)\ .
\end{equation*}
Since the action $\mathcal{V}_{l+1}$ does not survive the projection, the last indices in the constituent (\ref{eq::basisChiTilde}) of $\eta_q^{(l)}$ are still the same. Therefore, using again (\ref{eq::propertyEnergy1}), we have
\begin{equation}
\mathcal{P}_0(\mathcal{V}\eta_q^{(l)}\otimes I_1)=\beta_p^{(l)}\otimes I_1\ .    
\end{equation}
with $\beta_p^{l}\in Null(\mathcal{H}_0)$ and
\begin{equation}
    \beta_p^{(l)}=\mathcal{P}_0\mathcal{V}_1\eta_q^{(l)}+\mathcal{P}_0\mathcal{V}\rho_q^{(l)}\ .
\end{equation}
Now, if a solution for $\psi_p^{(l)}$ exists, because of Property \ref{property::GeneralConditionAlphaBeta}, the second part of Property 4', regarding the structure of $\psi_p^{(l)}$, follows. This concludes the proof.
\end{proof}

Note that we can easily find the coefficients $\eta^{i_1,i_2,\ldots,i_{l},i_{l+1}}_{j_1,j_2,\ldots,j_{l},i_{l+1}}$ of $\eta_q^{(l)}$. In fact we can see by induction that
\begin{equation}
   \eta^{i_1,i_2,\ldots,i_{l},i_{l+1}}_{j_1,j_2,\ldots,j_{l},i_{l+1}}=-\frac{(\omega^{j_1}-\omega^{i_1})}{E^{i_1,i_2,\ldots,i_{l},i_{l+1}}_{j_1,j_2,\ldots,j_{l},i_{l+1}}}\prod_{t=2}^le^{i(j_t-i_t)\phi}\left(\frac{1}{E^{i_1,i_2,\ldots,i_{t-1},i_{t}}_{j_1,j_2,\ldots,j_{t-1},i_{t}}}-\frac{1}{E^{i_1,i_2,\ldots,i_{t-1},j_{t}}_{j_1,j_2,\ldots,j_{t-1},j_{t}}}\right)\ . 
\end{equation}
From the fact that all the $i$s and $j$s are different up to the last site assures that all the energy in the formula are different from zero because of the conservation of total domain wall angle.
\newline
\\\textbf{Broken total domain wall angle}
\newline
\\We can now consider what happens when the total domain wall angle is not conserved. As we saw in Section \ref{sec::LocalityTotalDomainWallParity} resonance points appear when we consider chains of increasing length. From Property 4' we can expect that the first operators that can produce problems stems from the $\eta_q^{(k)}$ described above, as it contains the operators that act non-trivially on the largest number of sites. 

 Suppose therefore that $\theta$ is such that total domain wall angle is not conserved on chains of length $k+2$. Using Property \ref{property::Null} we know that the operators $\mathcal{P}_0(\mathcal{V}\psi_q^{(k)}\otimes I_1)$ are a linear combination of operators of the form
\begin{equation}
\label{eq::TempDWParity}
    |i_1,\mathbf{i}_{k-1},i_k,i_{k+2}\rangle|j_1,\mathbf{j}_{k-1},j_k,i_{k+2}\rangle\qquad i_1\neq j_1 \qquad E^{i_1,\mathbf{i}_{k-1},i_k,i_{k+2}}_{j_1,\mathbf{j}_{k-1},j_k,i_{k+2}}=0\ .
\end{equation}
We note that the existence of this type of terms is equivalent to saying that condition (134) does not hold,
otherwise we would have an identity at the end. Also note that these terms are bound to stem out from the $\eta_q^{(k)}$ of Property 4'\label{eq::AppendixconditionExistence}, and are invariably produced, as can be proved with a bit of effort. 

Using Property \ref{property::GeneralConditionAlphaBeta} we know that the solution $\psi^{(k)}_p$ has to live on a chain of length $k+1$, but for chains of smaller length the total domain wall angle is still conserved, which means, using Property \ref{property::Null}, that $\psi_p^{(k)}$ has to be the linear combination of vectors of the type
\begin{equation}
    |i_1,\mathbf{i}_{k-1},i_k\rangle|i_1,\mathbf{j}_{k-1}i_k\rangle\otimes I_1\ .
\end{equation}
Given the local structure of $\mathcal{V}$ this is impossible unless, $i_k=j_k$ in (\ref{eq::TempDWParity})\footnote{$\mathcal{V}$ in fact has to act through $\mathcal{V}_1$ to make $i_1\neq j_1$ in (\ref{eq::TempDWParity}).}. This means that
\begin{equation}
    E^{i_1,\mathbf{i}_{k-1},i_k,i_{k+2}}_{j_1,\mathbf{j}_{k-1},i_k,i_{k+2}}= E^{i_1,\mathbf{i}_{k-1},i_k}_{j_1,\mathbf{j}_{k-1},i_k}=0\ ,
\end{equation}
but this is not possible since by hypothesis for chains of length $k+1$ the total domain wall angle is conserved. This means that a local solution of  the expansion at order $k$ cannot exist at a resonance point appearing on chains of length $k+2$ when total domain wall angle symmetry is broken. This proves Property \ref{property::NonExistence}, that we repeat here for completeness
\newline
\\\textbf{Property 5.} \textit{
Suppose that $\theta$ is at a resonant point which does not conserve the total domain wall angle. Then the formal expansion for the zero mode can exist only up to an order compatible with the length of the chain at which the resonant point first appear.}

\section{Proof of the solution for $\psi_p^{(2)}$}
\label{appendix:ProofChip}

In this appendix we will provide a proof of the formula for the solution of $\psi_p^{(2)}$. To keep the exposition simple in the following we will consider $\phi=0$. The same arguments can be given also for the general case.

\subsection{Solution for $\psi^{(2)}$}

First of all from the proof of Property 4' we know that
\begin{equation}
    \begin{split}
        &\psi_q^{(1)}=\sum_{i_1\neq j_1,i_2} (\omega^{j_1}-\omega^{i_1})\frac{1}{E^{i_1,i_2}_{j_1,i_2}}|i_1,i_2\rangle|j_1,i_2\rangle\\
        &\psi_p^{(0)}=0\ .
    \end{split}
\end{equation}
and
\begin{equation*}
    \begin{split}
        \mathcal{V}\psi_q^{(1)}\otimes I_1&=-\sum_{i_1\neq j_1,i_2\neq j_2,i_3}(\omega^{j_1}-\omega^{i_1})\left(\frac{1}{E^{i_1,j_2}_{j_1,j_2}}-\frac{1}{E^{i_1,i_2}_{j_1,i_2}}\right)|i_1,i_2,i_3\rangle|j_1,j_2,i_3\rangle\\
        &-\sum_{i_1\neq j_1,i_2,i_3}(\omega^{j_1}-\omega^{i_1})\left(\frac{1}{E^{2i_1-j_1,i_2}_{\phantom{2i_1}j_1\phantom{-},i_2}}-\frac{1}{E^{\phantom{2j_1}i_1\phantom{-},i_2}_{2j_1-i_1,i_2}}\right)|i_1,i_2,i_3\rangle|j_1,i_2,i_3\rangle\ .
    \end{split}
\end{equation*}
We will now find the solution for $\psi^{(2)}=\psi_q^{(2)}+\psi_p^{(2)}$. As before we can solve for $\psi_q^{(2)}$ by simply inverting $\mathcal{H}_0$:
\begin{equation}
\label{eq::Appendixtemp3}
    \begin{split}
        \psi_q^{(2)}&=\sum_{i_1\neq j_1,i_2\neq j_2,i_3}\frac{\omega^{j_1}-\omega^{i_1}}{E^{i_1,i_2,i_3}_{j_1,j_2,i_3}}\left(\frac{1}{E^{i_1,j_2}_{j_1,j_2}}-\frac{1}{E^{i_1,i_2}_{j_1,i_2}}\right)|i_1,i_2,i_3\rangle|j_1,j_2,i_3\rangle\\
        &+\sum_{i_1\neq j_1,i_2}\frac{\omega^{j_1}-\omega^{i_1}}{E^{i_1,i_2}_{j_1,i_2}}\left(\frac{1}{E^{2i_1-j_1,j_2}_{\phantom{2i_1}j_1\phantom{-},j_2}}-\frac{1}{E^{\phantom{2j_1}i_1\phantom{-},i_2}_{2j_1-i_1,i_2}}\right)|i_1,i_2\rangle|j_1,i_2\rangle\otimes I_1\ .
    \end{split}        
\end{equation}
Note that (\ref{eq::Appendixtemp3}) follows the structure outlined in Property 4'. We are now in a position to prove that
\begin{equation}
    \psi_p^{(2)}=-\frac{1}{2}\mathcal{P}_0\mathcal{V}\left(\frac{\mathcal{Q}_0}{\mathcal{H}_0}\right)^2\mathcal{V}(\psi_p^{(0)}\otimes I_2)\ .
\end{equation}
First of all note that with this definition
\begin{equation*}
        \psi_p^{(2)}=-\frac{1}{2}\sum_{i_1\neq j_1,i_2}(\omega^{j_1}-\omega^{i_1})\left(\frac{1}{\left(E^{i_1,i_2}_{j_1,i_2}\right)^2}+\frac{1}{\left(E^{j_1,i_2}_{i_1,i_2}\right)^2}\right)|i_1,i_2\rangle|i_1,i_2\rangle\otimes I_1
\end{equation*}
and since $E^{j_1,i_2}_{i_1,i_2}=-E_{j_1,i_2}^{i_1,i_2}$ we have
\begin{equation}
        \psi_p^{(2)}=-\sum_{i_1\neq j_1,i_2}(\omega^{j_1}-\omega^{i_1})\frac{1}{\left(E^{i_1,i_2}_{j_1,i_2}\right)^2}|i_1,i_2\rangle|i_1,i_2\rangle\otimes I_1\ .
\end{equation}
The action of $\mathcal{V}$ on $\psi_p^{(2)}$ is now given by
\begin{equation}
\label{eq::Appendixtemp4}
    \begin{split}
        \mathcal{P}_0\mathcal{V}\psi_p^{(2)}=-\sum_{i_1\neq j_1,i_2\neq j_2,i_3}(\omega^{j_1}-\omega^{i_1})\left(\frac{1}{\left(E^{i_1,j_2}_{j_1,j_2}\right)^2}-\frac{1}{\left(E^{i_1,i_2}_{j_1,i_2}\right)^2}\right)|i_1,i_2,i_3\rangle|i_1,j_2,i_3\rangle\ .
    \end{split}
\end{equation}

We will now show that $\mathcal{P}_0(\mathcal{V}\psi_q^{(2)}\otimes I_1)$ yields the same expression. Since $\psi_q^{(2)}$ follows Property \ref{property::conditionExistence}' we know that
\begin{equation}
    \mathcal{P}_0(\mathcal{V}\psi_q^{(2)}\otimes I_1)= (\mathcal{P}_0\mathcal{V}_1\psi_q^{(2)})\otimes I_1
\end{equation}
This simplifies the problem, as we do not need to consider chains of length 4 and the problem stays confined on a chain of length 3~\footnote{Again, this is true only in the case where total domain wall angle is conserved. For chains of length 4 means $\theta\neq 0,\frac{\pi}{3},\frac{2\pi}{3},\arctan{\frac{\sqrt{3}}{5}}$.}. The action of $\mathcal{P}_0\mathcal{V}_1$ yelds
\begin{equation*}
    \begin{split}
        \mathcal{P}_0\mathcal{V}_1\psi_q^{(2)}&=\sum_{i_1\neq j_1,i_2\neq j_2,i_3}\frac{\omega^{j_1}-\omega^{i_1}}{E^{i_1,i_2,i_3}_{j_1,j_2,i_3}}\left(\frac{1}{E^{i_1,j_2}_{j_1,j_2}}-\frac{1}{E^{i_1,i_2}_{j_1,i_2}}\right)|i_1,i_2,i_3\rangle|i_1,j_2,i_3\rangle\\
        &-\sum_{i_1\neq j_1,i_2\neq j_2,i_3}\frac{\omega^{i_1}-\omega^{j_1}}{E^{j_1,i_2,i_3}_{i_1,j_2,i_3}}\left(\frac{1}{E^{j_1,j_2}_{i_1,j_2}}-\frac{1}{E^{j_1,i_2}_{i_1,i_2}}\right)|i_1,i_2,i_3\rangle|i_1,j_2,i_3\rangle
    \end{split}
\end{equation*}
and all the operators in the sums are such that $E^{i_1,i_2,i_3}_{i_1,j_2,i_3}=0$~\footnote{The terms with $i_2=j_2$ are identically zero, as can be checked by direct computation.}. Note that all the terms with $i_2=j_2$ cancel identically. Consider now the energies $E^{i_1,i_2,i_3}_{j_1,j_2,i_3}$, $E^{j_1,i_2,i_3}_{i_1,j_2,i_3}$ with $j_1\neq i_1$ as in the sums. We have
\begin{equation}
    \begin{split}
        &E^{i_1,i_2,i_3}_{j_1,j_2,i_3}=E^{i_1,i_2,i_3}_{i_1,j_2,i_3}+E^{i_1,j_2}_{j_1,j_2}=E^{i_1,j_2}_{j_1,j_2}\\
        &E^{j_1,i_2,i_3}_{i_1,j_2,i_3}=E^{i_1,i_2,i_3}_{i_1,j_2,i_3}+E^{j_1,i_2}_{j_1,i_2}=E^{j_1,i_2}_{i_1,i_2}\ ,
    \end{split}
\end{equation}
where we used that $E^{i_1,i_2,i_3}_{i_1,j_2,i_3}=0$. Hence we get 
\begin{equation*}
    \begin{split}
        \mathcal{P}_0\mathcal{V}\psi_q^{(2)}&=\sum_{i_1\neq j_1,i_2\neq j_2,i_3}(\omega^{j_1}-\omega^{i_1})\left(\frac{1}{(E^{i_1,j_2}_{j_1,j_2})^2}-\frac{1}{E^{i_1,j_2}_{j_1,j_2}E^{i_1,i_2}_{j_1,i_2}}\right)|i_1,i_2,i_3\rangle|j_1,j_2,i_3\rangle\\
        &-\sum_{i_1\neq j_1,i_2\neq j_2,i_3}(\omega^{i_1}-\omega^{j_1})\left(\frac{1}{E^{j_1,i_2}_{i_1,i_2}E^{j_1,j_2}_{i_1,j_2}}-\frac{1}{(E^{j_1,i_2}_{i_1,i_2})^2}\right)|i_1,i_2,i_3\rangle|j_1,j_2,i_3\rangle\ .
    \end{split}
\end{equation*}
Now, since $E^{j_1,i_2}_{i_1,i_2}=-E_{j_1,i_2}^{i_1,i_2}$ and $E^{j_1,j_2}_{i_1,j_2}=-E_{j_1,j_2}^{i_1,j_2}$, we are left with 
\begin{equation}
        \mathcal{P}_0\mathcal{V}\psi_q^{(2)}=\sum_{i_1\neq j_1,i_2\neq j_2,i_3}(\omega^{j_1}-\omega^{i_1})\left(\frac{1}{(E^{i_1,j_2}_{j_1,j_2})^2}-\frac{1}{(E^{j_1,i_2}_{i_1,i_2})^2}\right)|i_1,i_2,i_3\rangle|j_1,j_2,i_3\rangle\ ,
\end{equation}
this is the same as (\ref{eq::Appendixtemp4}). Therefore, to summarize we have checked that a solution of
\begin{equation}
\label{eq::Appendixtemp6}
    \mathcal{P}_0\mathcal{V}(\psi_p^{(2)}\otimes I_1)=-\mathcal{P}_0\mathcal{V}(\psi_q^{(2)}\otimes I_1)\ ,
\end{equation}
exists and is given by
\begin{equation}
    \psi_p^{(2)}=-\frac{1}{2}\mathcal{P}_0\mathcal{V}\left(\frac{\mathcal{Q}_0}{\mathcal{H}_0}\right)^2\mathcal{V}(\psi_p^{(0)}\otimes I_2)\ .
\end{equation}

\subsubsection{Normalization}

As we noted in the main text this solution is not unique and we can consider any linear combination of the form
\begin{equation}
    \psi^{(2)}_p+\xi_2\sigma_1 
\end{equation}
and we would still get a solution of (\ref{eq::Appendixtemp6}), for any $\xi_2\in\mathbb{C}$. As we saw in this Appendix and in Section \ref{sec::Normalization} the value of $\xi_2$ can be determined from the condition 
\begin{equation}
    (\psi^2)^{(2)}=(\psi^\dagger)^{(2)}\ .
\end{equation}
Computing the value of $\xi_2$ directly is rather bothersome and the algebra is quite involved, but we proved in Section \ref{sec::Normalization} that we can always find a $\xi_2$ such that this condition is satisfied. Using Mathematica, for $\theta=\frac{\pi}{6}$, we get\footnote{Also in the case of $\theta=\frac{\pi}{4}$ we get
\begin{equation*}
    \xi_2=0\ .
\end{equation*}
Note however that in general these constants are different from 0 and depend on $\theta$. For example in the case $\theta=\frac{\pi}{6}$ we get
\begin{equation*}
    \xi_3=\frac{i}{6}\ ,
\end{equation*}
while in the case of $\theta=\frac{\pi}{4}$ we have
\begin{equation*}
    \xi_3=\frac{i}{3\sqrt{2}}
\end{equation*}
} 
\begin{equation}
    \xi_2=0\ .
\end{equation}
The same analysis conducted with symbolical computation also shows that with this choice of $\xi_2$ we get 
\begin{equation}
    (\psi^3)^{(2)}=0\ .
\end{equation}
So, overall
\begin{equation}
    \psi^{(2)}=\left(\left(\frac{\mathcal{Q}_0}{\mathcal{H}_0}\right)\mathcal{V}\left(\frac{\mathcal{Q}_0}{\mathcal{H}_0}\right)\mathcal{V}+\frac{1}{2}\mathcal{P}_0\mathcal{V}\left(\frac{\mathcal{Q}_0}{\mathcal{H}_0}\right)^2\mathcal{V}\right)\psi^{(0)}\ .
\end{equation}

\section{Coefficients for the solution $\psi_p^{(5)}$}
\label{appendix::GammasForPsip5}

In this appendix we will display, for completeness, the non-zero $\Gamma_{i_1,i_2,i_3,i_4}$ introduced in Section \ref{sec::GeneralFormOfTheSolution}, which appear in $\psi_p^{(5)}$ as found through the use of Mathematica. We have
\begin{equation}
\begin{array}{c c c c}
        \Gamma_{0,0,4,1}=-\frac{3}{35}&\Gamma_{0,0,3,2}=-\frac{2}{35}&\Gamma_{0,3,0,2}=-\frac{3}{14}&\Gamma_{0,1,1,3}=\phantom{-}\frac{1}{35}\vspace{0.2cm}\\
        \Gamma_{0,1,3,1}=\phantom{-}\frac{6}{35}&\Gamma_{1,0,3,1}=\phantom{-}\frac{1}{5}&\Gamma_{0,3,1,1}=\phantom{-}\frac{3}{7}&\Gamma_{0,1,2,2}=\phantom{-}\frac{3}{35}\vspace{0.2cm}\\
        \Gamma_{1,0,2,2}=\phantom{-}\frac{1}{5}&\Gamma_{0,2,1,2}=\phantom{-}\frac{1}{7}&\Gamma_{2,0,1,2}=\phantom{-}\frac{7}{20}&\Gamma_{1,2,0,2}=\phantom{-}\frac{3}{10}\vspace{0.2cm}\\
        \Gamma_{2,1,0,2}=\phantom{-}\frac{2}{5}&\Gamma_{0,2,2,1}=\phantom{-}\frac{2}{7}&\Gamma_{2,0,2,1}=\phantom{-}\frac{9}{20}&\Gamma_{1,1,1,2}=-\frac{1}{5}\vspace{0.2cm}\\
        \Gamma_{1,1,2,1}=-\frac{2}{5}&\Gamma_{1,2,1,1}=-\frac{3}{5}&\Gamma_{2,1,1,1}=-\frac{4}{5}\ .&
\end{array}
\end{equation}

\end{document}